\documentclass[11pt,a4paper]{article}

\usepackage{amsmath}
\usepackage{amssymb}
\usepackage{graphicx}
\usepackage[svgnames]{xcolor}
\usepackage{float}
\usepackage{microtype}
\usepackage{authblk}

\usepackage[numbers,sort&compress]{natbib}
\usepackage[
    colorlinks=true,
    linkcolor=blue,
    citecolor=blue,
    urlcolor=blue
]{hyperref}

\title{\textbf{Beyond de Sitter: Longitudinal-Mode Instabilities and Spectral Evolution of Massive Vector Fields with Non-Minimal Curvature Coupling during Inflation}}

\author[1,2]{E Momeni\thanks{Email:
\href{mailto:ehsan.momeni@ut.ac.ir}{ehsan.momeni@ut.ac.ir},
\href{mailto:ehsan.momeni@studenti.unipd.it}{ehsan.momeni@studenti.unipd.it}}}

\author[2]{F Shojai\thanks{Email:
\href{mailto:fshojai@ut.ac.ir}{fshojai@ut.ac.ir}}}

\author[3]{H Khodabakhshi\thanks{Email:
\href{mailto:h_khodabakhshi@tju.edu.cn}{h\_khodabakhshi@tju.edu.cn}}}

\affil[1]{Department of Physics and Astronomy ``Galileo Galilei'', University of Padova, Padova, Italy}

\affil[2]{Department of Physics, University of Tehran, Tehran, Iran}

\affil[3]{Center for Joint Quantum Studies and Department of Physics, School of Science, Tianjin University, Tianjin 300350, China}

\date{}

\begin{document}

\maketitle

\begin{abstract}
\noindent
We investigate the inflationary dynamics and spectral energy densities of a
massive spectator vector field non-minimally coupled to the Ricci scalar.
We compare exact de Sitter evolution with a constant-slow-roll
quasi-de Sitter background and follow the transverse and longitudinal
modes through an idealized instantaneous transition to radiation
domination. The modes are initialized deep inside the horizon using
second-order WKB Bunch--Davies initial conditions, and the mode
equations are solved numerically for representative comoving momenta
and vector masses.

\noindent
The non-minimal coupling modifies the effective mass, canonical
normalization, and effective-mass-crossing time, while the
horizon-crossing condition remains fixed by the prescribed background
in the spectator-field approximation.
The mass-crossing spectral energy densities exhibit
distinct responses in the transverse and longitudinal
sectors as the non-minimal coupling parameter $\xi$
increases. For the transverse modes, the spectral energy
density changes only weakly, showing a slight monotonic
increase over the investigated range
$0\leq\xi\leq10^{-6}$. In contrast, the longitudinal
spectral energy density decreases monotonically and
more substantially, with an overall suppression of
approximately $21\%$ between $\xi=0$ and
$\xi=10^{-6}$ in both dS and qdS backgrounds.
The qdS longitudinal values remain slightly higher
than their dS counterparts, whereas the transverse
results are nearly indistinguishable in the two
backgrounds. Despite this suppression, the
longitudinal contribution remains dominant at the
mass-crossing time, exceeding the transverse
contribution by approximately three orders of
magnitude.

\noindent
We also derive an all-momentum condition for a non-negative
longitudinal canonical frequency in exact de Sitter space. For the
representative mass considered, the required coupling lies outside the
constant-coefficient heavy-scalar matching regime; consequently,
large-coupling solutions should be interpreted as formal rather than
controlled effective-theory predictions. Our results clarify the
separate roles of kinetic positivity, canonical-frequency signs, bare
spectral energy densities, and the domain of validity of the effective
description.
\end{abstract}


\section{Introduction}

The inflationary production of massive vector fields has been extensively
studied. In particular, the longitudinal polarization exhibits dynamics that differ from those of the two transverse polarizations, while nonminimal
gravitational couplings can modify the effective mass and canonical
normalization of the field. Previous works have examined the evolution of
massive vector modes in inflationary backgrounds and their subsequent
cosmological evolution, including gravitational particle production and the
associated spectral energy densities
\cite{Capanelli2024,GrzadkowskiSocha2025,OzsoyTasinato2024}. These studies
provide the theoretical background for the present analysis.

\noindent
Inflation provides a compelling framework for describing the very early
Universe~\cite{Baumann2009,Baumann2012,Matarrese2019,Peacock1999,LiddleLyth2000,Riotto2003},
offering not only a mechanism for accelerated expansion but also a source of
the primordial perturbations that later seed the anisotropies of the cosmic
microwave background~\cite{Hu2002,Bartolo2006,Aghanim2020}
(see also~\cite{Mollerach1995}) and the formation of large-scale structure.
Although scalar fields are the simplest and most widely studied candidates
for driving inflation, higher-spin fields---and in particular massive vector fields---have attracted increasing attention as possible extensions of this framework. Massive vector fields have been studied in several inflationary
and post-inflationary contexts, including vector
inflation~\cite{Golovnev2008}, vector curvaton
scenarios~\cite{Dimopoulos2008}, and the production of vector
dark matter during inflation~\cite{Graham2016}. More generally, vector
fields have been investigated in a variety of inflationary settings, as well
as for their potential signatures in primordial perturbations and other
cosmological observables~\cite{Dimastrogiovanni2010,Dimopoulos2009}. Non-minimal couplings
between vector fields and the curvature scalar have also been considered in
this broader context. Such couplings modify the dynamics of the vector modes and, in particular, require a careful analysis of the longitudinal sector and its stability.

\noindent
Unlike scalar degrees of freedom, massive vector fields possess both
transverse and longitudinal polarization modes, whose cosmological
evolution can differ significantly in an expanding background. In
particular, the longitudinal mode is known to be more sensitive to
potential instabilities, such as ghost-like or tachyonic
behavior~\cite{Himmetoglu2009a,Himmetoglu2009Ghost,EspositoFarese2010,Jimenez2012},
whereas minimally coupled vector fields are typically strongly suppressed
during inflation. Massive vector fields are particularly interesting because
their longitudinal polarization is absent in Maxwell theory but becomes
dynamical once the vector field acquires a mass. In an expanding Universe,
this longitudinal degree of freedom can behave very differently from the
transverse polarizations. Its kinetic normalization, effective frequency,
and stability properties depend sensitively on the ratio between the
physical momentum, the vector mass, and the expansion rate.
Since the vector field considered here is a spectator, its behavior also
depends strongly on the hierarchy between its mass and the Hubble rate
during inflation~\cite{Capanelli2024,GrzadkowskiSocha2025,OzsoyTasinato2024,Graham2016}, and gravitational production of other
spectator-type fields in the inflationary universe has long been
studied~\cite{Ema2019}. Ghost-free extensions of the Proca action and
their cosmology have been studied
extensively~\cite{Heisenberg2014,DeFelice2016Cosmology,DeFelice2016Effective,Heisenberg2016,Heisenberg2017,Heisenberg2018},
and vector fields in generalized geometries have been used to drive
inflation~\cite{Jimenez2016Distortion}.
One way to
improve the cosmological viability of such models is to introduce a
non-minimal coupling between the vector field and the Ricci scalar,
thereby modifying the field's effective mass and its dependence on the
background geometry.
Non-minimal curvature couplings of this type are familiar from
scalar-tensor models such as Brans--Dicke
inflation~\cite{Starobinsky1995}.
This coupling is also motivated by the
fact that in a time-dependent cosmological background the Ricci scalar
changes between different epochs. During inflation, $R$ is approximately
constant and positive, while in the RD era of a spatially
flat FLRW universe it vanishes. Therefore, the non-minimal coupling
naturally provides an epoch-dependent effective mass. This makes it
possible to study how vector-field production, stability, and energy
density depend on the transition from inflation to RD.

\noindent
The dynamics of these vector fluctuations are strongly influenced by
their coupling to the spacetime curvature. In this work, we consider a
massive vector field non-minimally coupled to the Ricci scalar $R$,
leading to an effective mass
\begin{equation}
    M^2(\tau) = m^2 + \xi R(\tau),
\end{equation}
where $\xi$ is a dimensionless coupling parameter. This interaction
provides a direct link between the background geometry and the evolution
of the vector modes, modifying the mass-crossing conditions and the
stability properties of the longitudinal sector. Such couplings have
appeared in vector inflation and vector-curvaton
constructions~\cite{Golovnev2008,Dimopoulos2008}, but they also require a
careful analysis of the stability of the longitudinal
sector~\cite{LebedevYoon2026}.

\noindent
A key objective of our study is to evaluate the validity of the exact
de Sitter (dS) approximation, which is often used for its mathematical
simplicity. A realistic inflationary background is only qdS, characterized by
a small but non-zero slow-roll parameter~\cite{Baumann2009,LiddleLyth2000,Liddle1994}
\begin{equation}
\epsilon \equiv -\frac{\dot{H}}{H^2}.
\label{eq:slow_roll_parameter}
\end{equation}

\noindent
The standard polarization decomposition and canonical frequencies are
presented in \cite{Capanelli2024}, while
\cite{GrzadkowskiSocha2025} derives the nonminimal stress-energy tensor
and investigates the corresponding spectral energy densities. We adopt the
relevant established results and specify the equations required for our
conventions and numerical calculations, without claiming these derivations
as new. We also emphasize that the Ricci-scalar coupling considered here is
distinct from theories involving an independent coupling to the Ricci
tensor.

\noindent
A further issue concerns the effective-field-theory interpretation of the curvature coupling. Lebedev and Yoon \cite{LebedevYoon2026} have shown
that a constant non-minimal coupling of a massive gauge vector requires a careful ultraviolet interpretation, including the
restoration of gauge invariance in the massless limit and the energy range over which heavy degrees of freedom can
consistently be integrated out. In explicit gauge-invariant completions, the induced coupling is suppressed by the vector
mass squared relative to an appropriate heavy scale. The validity of the resulting constant-coefficient effective description
must therefore be checked against both the background curvature and the physical momenta involved in the mode evolution.
Positivity of the reduced longitudinal kinetic term is necessary for the absence of a quadratic ghost, but it does not establish
ultraviolet validity.

\noindent
Within this framework, we compare exact dS and constant-slow-roll
evolution at common times and at coupling-dependent effective-mass-crossing
events. We investigate the longitudinal canonical-frequency threshold and
the leading infrared coefficient of the spectral energy density. These
diagnostics address distinct aspects of the mode evolution: comparisons at
different mass-crossing times do not constitute common-time comparisons,
while negative contributions to the bare spectral energy do not imply
negative particle occupation numbers. We retain large-coupling examples
only as formal solutions when they lie outside the applicable matching
regime. Accordingly, we do not infer a relic abundance, observational
bound, or renormalized total stress-energy from the results.

\noindent
In this paper, we numerically analyze the evolution of the spectral
energy densities for the transverse and longitudinal modes of a
non-minimally coupled vector field in dS, qdS, and RD backgrounds.
Specifically, we quantify the influence of the non-minimal coupling
parameter $\xi$ and slow-roll corrections on the suppression or
enhancement of the spectral energy densities of both polarization modes
and on the timing of the effective-mass crossing during inflation.

\noindent
The paper is organized as follows. Section~2 introduces the
backgrounds considered in this work, namely the dS, qdS, and RD eras,
together with the instantaneous inflation--radiation transition.
Section~3 presents the test massive vector field in an FLRW background:
it derives the action, mode equations, and spectral energy densities for the non-minimally coupled case, reduces them to the minimally coupled
case, and discusses the ghost and tachyonic stability conditions. Section~4 presents the numerical results for the spectral
energy densities of the transverse and longitudinal modes, comparing the dS and qdS backgrounds, and analyzes the effective frequency of the longitudinal mode. Finally, Section~5 summarizes the results and presents the conclusions. The appendices A, B, C, and D close the paper.

\clearpage
\section{Background Evolution}

\noindent
We consider a spatially flat FLRW spacetime~\cite{Matarrese2019,Peacock1999,Hartle2003} described by
\begin{equation}
ds^2=a^2(\tau)
\left(
-d\tau^2+d\mathbf{x}^{2}
\right),
\label{eq:FLRW_metric}
\end{equation}
where $a(\tau)$ is the scale factor and $\tau$ denotes the conformal
time. We investigate the evolution of massive vector fields during
inflation, considering both dS and qdS
backgrounds followed by the RD era.

\subsection{dS and qdS Backgrounds}

\noindent
In an exact dS background, the Hubble parameter is constant,
\begin{equation}
H=H_{\rm end},
\label{eq:dS_Hubble}
\end{equation}
and the scale factor evolves as
\begin{equation}
a_{\rm dS}(\tau)
=
-\frac{1}{H_{\rm end}\tau}.
\label{eq:dS_scale_factor}
\end{equation}

\noindent
To describe deviations from exact dS expansion, we consider a qdS
background characterized by the slow-roll parameter of \eqref{eq:slow_roll_parameter}
where $0<\epsilon\ll1$ during inflation. 
Assuming a constant slow-roll parameter~\cite{Liddle1994}, the Hubble parameter evolves with the number of e-folds $N$ as
\begin{equation}
H(N)=H_{\rm end}e^{-\epsilon N}.
\label{eq:qdS_Hubble}
\end{equation}

\noindent
Defining
\begin{equation}
\alpha\equiv1-\epsilon,
\label{eq:alpha_definition}
\end{equation}
the scale factor in conformal time is given by
\begin{equation}
a_{\rm qdS}(\tau)
=
\left(
-\frac{1}{\alpha H_{\rm end}\tau}
\right)^{\frac{1}{\alpha}}.
\label{eq:qdS_scale_factor}
\end{equation}

\subsection{Inflation-Radiation Transition}

\noindent
Following inflation, we assume an instantaneous transition to the RD era. This approximation introduces a discontinuity in the Ricci scalar. As discussed in Section~4, the non-minimal coupling links the vector field's effective mass to the background curvature; consequently, the curvature-dependent contribution to the mass vanishes abruptly at the onset of RD. Therefore, any rapid features observed in the spectra near $N=0$ should be interpreted as artifacts of this instantaneous-transition approximation, which would likely be mitigated in a more realistic smooth reheating scenario. The scale factor and its first derivative are matched continuously at the transition time. We set the normalization by
imposing

\begin{equation}
a(\tau_{\rm end})=1 ,
\end{equation}
With this choice, the scale factor evolution can be written as

\begin{equation}
a_{\mathrm{dS}\rightarrow\mathrm{RD}}(\tau)=
\begin{cases}
-\dfrac{1}{H_{\mathrm{end}}\tau},
&
\tau < -\dfrac{1}{H_{\mathrm{end}}},
\\[10pt]
H_{\mathrm{end}}\tau + 2,
&
\tau \geq -\dfrac{1}{H_{\mathrm{end}}}.
\end{cases}
\label{eq:scale_factor_dS_RD}
\end{equation}

\begin{equation}
a_{\mathrm{qdS}\rightarrow\mathrm{RD}}(\tau)=
\begin{cases}
\left(
-\dfrac{1}{\alpha H_{\mathrm{end}}\tau}
\right)^{\frac{1}{\alpha}},
&
\tau < -\dfrac{1}{\alpha H_{\mathrm{end}}},
\\[10pt]
H_{\mathrm{end}}\tau + \dfrac{1+\alpha}{\alpha},
&
\tau \geq -\dfrac{1}{\alpha H_{\mathrm{end}}}.
\end{cases}
\label{eq:scale_factor_qdS_RD}
\end{equation}

\noindent
We extend the e-fold variable through the transition by defining $N=\ln a$, with $a_{\mathrm{end}}=1$. Hence, $N<0$ during inflation, $N=0$ at the transition, and $N>0$ during the RD era.

\begin{equation}
	\tau_{\mathrm{dS}\rightarrow \mathrm{RD}}(N)=
	\begin{cases}
		-\dfrac{e^{-N}}{H_{\mathrm{end}}},
		& N < 0, \\[8pt]
		
		\dfrac{e^{N}}{H_{\mathrm{end}}}
		-\dfrac{2}{H_{\mathrm{end}}},
		& N \geq 0.
	\end{cases}
	\label{eq:tau_N}
\end{equation}

\begin{equation}
	\tau_{\mathrm{qdS}\rightarrow \mathrm{RD}}(N)=
	\begin{cases}
		-\dfrac{e^{-\alpha N}}{\alpha H_{\mathrm{end}}},
		& N < 0, \\[8pt]
		
		\dfrac{e^{N}}{H_{\mathrm{end}}}
		-\dfrac{1+\alpha}{\alpha H_{\mathrm{end}}},
		& N \geq 0.
	\end{cases}
	\label{eq:tau_N_phases}
\end{equation}
\noindent
Therefore, the change of variables between conformal time $\tau$ and the e-folding number $N$ in the dS, qdS, and RD phases depends on the matching prescription. Consequently, the derivative operators transform accordingly in each regime.

\begin{equation}
	\frac{d}{d\tau}
	= H_{\text{end}} e^{\gamma N}\frac{d}{dN},
	\qquad
	\frac{d^2}{d\tau^2}
	= H_{\text{end}}^2 e^{2\gamma N}
	\left(
	\frac{d^2}{dN^2}
	+ \gamma \frac{d}{dN}
	\right)
	\label{eq:tau_to_N_derivatives}
\end{equation}
where 
\begin{equation}
	\gamma =
	\begin{cases}
		1, & \text{(dS)}, \\[6pt]
		\alpha, & \text{ (qdS)}, \\[6pt]
		-1, & \text{(RD)}.
	\end{cases}
	\label{eq:gamma_definition}
\end{equation}

\subsubsection{Energy Density in the dS, qdS, and RD Regimes}

\noindent
For a spatially flat universe with vanishing cosmological constant,
the first Friedmann equation~\cite{Peacock1999,Hartle2003} gives
\begin{equation}
    \rho = 3 M_{\mathrm{Pl}}^2 H^2,
    \qquad
    M_{\mathrm{Pl}} \equiv (8\pi G)^{-1/2}.
    \label{eq:first_friedmann_equation}
\end{equation}
We model the transition from inflation to radiation domination as
instantaneous and impose continuity of the scale factor and the Hubble
parameter at $\tau=\tau_{\mathrm{end}}$. Setting
$a_{\mathrm{end}}=1$, this implies
\begin{equation}
    H_{\mathrm{RD}}(\tau_{\mathrm{end}})
    =
    H_{\mathrm{end}}.
    \label{eq:hubble_matching}
\end{equation}
Since the background energy density scales as $\rho_{\mathrm{RD}}\propto
a^{-4}$ during radiation domination, the Hubble parameter satisfies
\begin{equation}
    H_{\mathrm{RD}}^2(\tau)
    =
    H_{\mathrm{end}}^2 a^{-4}(\tau).
    \label{eq:hubble_RD}
\end{equation}

\noindent
The background energy density for the dS-to-RD transition is therefore
\begin{equation}
    \rho_{\mathrm{dS}\rightarrow\mathrm{RD}}(\tau)
    =
    3 M_{\mathrm{Pl}}^2 H_{\mathrm{end}}^2
    \begin{cases}
        1,
        & \tau < \tau_{\mathrm{end}}, \\[8pt]
        a^{-4}(\tau),
        & \tau \geq \tau_{\mathrm{end}}.
    \end{cases}
    \label{eq:energy_density_dS_RD}
\end{equation}
For a qdS-to-RD transition with constant slow-roll parameter
$\epsilon$, one has $H\propto a^{-\epsilon}$ during qdS inflation.
The corresponding energy density is
\begin{equation}
    \rho_{\mathrm{qdS}\rightarrow\mathrm{RD}}(\tau)
    =
    3 M_{\mathrm{Pl}}^2 H_{\mathrm{end}}^2
    \begin{cases}
        a^{-2\epsilon}(\tau),
        & \tau < \tau_{\mathrm{end}}, \\[8pt]
        a^{-4}(\tau),
        & \tau \geq \tau_{\mathrm{end}}.
    \end{cases}
    \label{eq:energy_density_qdS_RD}
\end{equation}

\section{Test Massive Vector Field in an FLRW Background}
We consider a spatially flat FLRW spacetime written in conformal time, as given in Eq.~\eqref{eq:FLRW_metric}. We restrict the coupling parameter to the range
\begin{equation}
    0 \leq \xi \leq 1 \times 10^{-6}.
    \label{eq:xi_range}
\end{equation}
The limit $\xi=0$ corresponds to the minimally coupled Proca field.

\subsection{Non-Minimal Curvature Coupling: $\xi \neq 0$}
\label{subsec:nonminimal_curvature_coupling}

In this subsection, we evaluate the action and the corresponding energy densities of the transverse and longitudinal modes of the massive vector field, considering a non-minimal coupling to curvature with $\xi\neq0$.

\subsubsection{Action, effective mass, and field equations}

The action is given by
\begin{equation}
    S
    =
    \int d^4x\,\sqrt{-g}
    \left[
        -\frac{1}{4} F_{\mu\nu}F^{\mu\nu}
        -\frac{1}{2} m^{2} A_\mu A^\mu
        -\frac{1}{2} \xi R A_\mu A^\mu
    \right]
    \label{eq:nonminimal_vector_action}
\end{equation}
where $F_{\mu\nu}= \partial_\mu A_\nu-\partial_\nu A_\mu$, $m$ is the bare vector mass, $\xi$ is a dimensionless non-minimal coupling parameter in four-dimensional spacetime, and $R$ is the Ricci scalar of the background spacetime.

\subsubsection{Domain of Validity and Notation}

\noindent
For the coupling samples adopted in the numerical analysis,
the curvature-expansion parameter provides an additional
consistency check. For
$m/H_{\rm end}=10^{-2}$ and $\epsilon=10^{-3}$, the largest
sample, $\xi=10^{-6}$, gives
\begin{equation}
    3(2-\epsilon)\frac{|\xi|H^2}{m^2}
    \simeq 0.0614
    \qquad (N=-17).
\end{equation}
Thus, this parameter remains below unity at this reference
time. Its value should nevertheless be assessed over the
full integration interval, since the controlled-expansion
condition requires it to be much smaller than unity.
Moreover, satisfying the curvature-expansion condition
alone does not establish the validity of the constant-
coupling effective description at all physical momenta.
In particular, integrating out the heavy scalar also
requires $p\ll m_s$. The form-factor considerations
discussed in Ref.~\cite{LebedevYoon2026} must therefore
also be taken into account when assessing the high physical
momenta used to initialize the numerical modes.

\noindent
For a spatially flat FLRW background written in conformal time, the Ricci scalar is
\begin{equation}
    R
    =
    6\,\frac{a''}{a^3}.
    \label{eq:ricci_scalar}
\end{equation}
Consequently, according to Eqs.~\eqref{eq:scale_factor_dS_RD} and \eqref{eq:scale_factor_qdS_RD}, the effective mass in the dS, qdS, and RD regimes can be written as
\begin{equation}
    M^2(\tau) = m^2 + 6 \xi (1 + \gamma) H_{\rm end}^2 \, a^{2(\mu - 1)},
    \label{eq:Meff_piecewise}
\end{equation}
where
\begin{equation}
    \mu =
    \begin{cases}
        1,      & \text{dS/RD}, \\[4pt]
        \alpha, & \text{qdS}.
    \end{cases}
    \label{eq:mu_definition}
\end{equation}
together with
	\begin{equation}
		-\frac{M^{2}}{2}A_{\mu}A^{\mu}
		=
		\frac{M^{2}}{2a^{2}}
		\left(A_{0}^{2}-|A|^{2}\right),
	\end{equation}
	we obtain the full action:
	\begin{equation}
		S=\frac{1}{2}\int d^{4}x
		\left[
		|A' - \nabla A_{0}|^{2}
		- |\nabla \times A|^{2}
		+ M^{2}a^{2}(A_{0}^{2}-|A|^{2})
		\right].
	\end{equation}
	We Fourier transform:
	\begin{equation}
		A_{\mu}(\tau, \mathbf{x}) = \int \frac{d^{3}k}{(2\pi)^{3/2}}\, e^{i\mathbf{k}\cdot\mathbf{x}}\, A_{\mu}(\tau,\mathbf{k}).
	\end{equation}

\noindent
We decompose the vector field into the following:
\begin{equation}
    A_\mu = \{A_0,\mathbf{A}\},
    \qquad
    \mathbf{A} = \mathbf{A}_T + A_L \hat{\mathbf{k}},
\end{equation}
where $\mathbf{k}\cdot\mathbf{A}_T=0$ and thus
$\mathbf{k}\cdot\mathbf{A}=kA_L$. We therefore see that the
longitudinal component $A_L$ and the transverse component
$\mathbf{A}_T$ contain, respectively, one and two degrees of
freedom. We also note that
$A_T(-\mathbf{k})=A_T^*(\mathbf{k})$, while
$A_L(-\mathbf{k})=-A_L^*(\mathbf{k})$.

\noindent
With this decomposition, the action in Fourier space reads as follows:
\begin{equation}
\begin{aligned}
S = \frac{1}{2}\int d\tau\,d^3k\,
\Bigg[
&|A_T'|^2
-(k^2+M^2a^2)|A_T|^2
-M^2a^2|A_L|^2
\\
&+\frac{M^2a^2}{k^2+M^2a^2}|A_L'|^2
+(k^2+M^2a^2)
\left|
\frac{ikA_L'}{k^2+M^2a^2}+A_0
\right|^2
\Bigg].
\end{aligned}
\end{equation}
Integrating $A_0$ results in the following:
\begin{equation}
  A_0 = -\frac{ikA_L'}{k^2+M^2a^2},
  \qquad
  k^2+M^2a^2\neq 0.
  \label{eq:A0_constraint}
\end{equation}
which eliminates the last term from the action.

\noindent
The resulting action is therefore the sum of two contributions,
one for the transverse components and one for the longitudinal
component.
\begin{equation}
    S = S_T + S_L,
\end{equation}
with
\begin{equation}
    S_T = \frac{1}{2}\int d\tau\,d^3k
    \sum_{i=1}^{2}
    \left[
    |A_{T,i}'|^2
    -(k^2+M^2a^2)|A_{T,i}|^2
    \right].
    \label{eq:transverse_reduced_action}
\end{equation}

\begin{equation}
    S_L = \frac{1}{2}\int d\tau\,d^3k
    \left[
    \frac{M^2a^2}{k^2+M^2a^2}|A_L'|^2
    -M^2a^2|A_L|^2
    \right].
    \label{eq:longitudinal_reduced_action}
\end{equation}
where we define
\begin{equation}
    K_L \equiv \frac{M^2 a^2}{k^2 + M^2 a^2}.
    \label{eq:K_L}
\end{equation}

\noindent
For $M^2>0$, the positivity of $K_L$ guarantees a positive kinetic
normalization for the longitudinal mode.

\noindent
Varying the reduced actions in
Eqs.~\eqref{eq:transverse_reduced_action} and
\eqref{eq:longitudinal_reduced_action} with respect to the corresponding
fields gives the equations of motion. For the transverse modes,
\begin{equation}
A_{T,i}''+
\left(k^2+M^2a^2\right)A_{T,i}=0.
\label{eq:ATT_eom}
\end{equation}
For the longitudinal mode, the equation of motion is
\begin{equation}
A_L''
+
\frac{2 k^2}{k^2+M^2a^2}
\left(
\frac{M'}{M}+\frac{a'}{a}
\right)A_L'
+
\left(k^2+M^2a^2\right)A_L=0.
\label{eq:ALL_eom}
\end{equation}

\subsubsection{Spectral Energy Density}

The action associated with the nonminimal coupling term is
\begin{equation}
S_{\xi}
=
-\frac{1}{2}
\int d^4x\,
\sqrt{-g}\,
\xi R A_\mu A^\mu.
\label{eq:action_xi}
\end{equation}
Consequently, the energy--momentum tensor receives
contributions from both the Proca sector and the
nonminimal coupling term, and can be decomposed as
\begin{equation}
T_{\mu\nu}
=
T_{\mu\nu}^{\rm(Proca)}
+
T_{\mu\nu}^{(\xi)}.
\label{eq:Tmunu_decomp}
\end{equation}
The energy density is obtained from
\begin{equation}
\rho
=
-g^{00}T_{00}
=
\frac{T_{00}}{a^2}.
\label{eq:rho_compact}
\end{equation}
After quantization and evaluation of the vacuum
expectation value, the total bare energy density can
be written as an integral over momentum modes,
\begin{equation}
\left\langle\rho\right\rangle
=
\frac{1}{2a^4}
\int\frac{d^3k}{(2\pi)^3}
\left[
\rho_T(\mathbf{k})+\rho_L(\mathbf{k})
\right].
\label{eq:rho_TL_integral}
\end{equation}
Here, $\rho_T(\mathbf{k})$ denotes the combined
contribution of the two transverse polarizations,
whereas $\rho_L(\mathbf{k})$ denotes the contribution
of the single longitudinal polarization. The overall
normalization follows from the definitions of the
corresponding mode functions and spectral kernels.
The resulting quantities are bare spectral energy
densities and are not, by themselves, renormalized
physical energy densities. \cite{GrzadkowskiSocha2025}

\subsubsection*{Transverse Mode}

To separate the transverse and longitudinal sectors, we decompose the spatial vector field in Fourier space as \(A_i=A_i^{(T)}+ik_iA_L\), where \(k_iA_i^{(T)}=0\). In the transverse sector, the longitudinal amplitude vanishes, \(A_L=0\). Moreover, the temporal component \(A_0\) is nondynamical and is determined by the temporal constraint equation. For transverse modes with nonzero comoving momentum, this constraint yields \(A_0=0\), provided the corresponding effective mass coefficient is nonzero. Therefore,
\begin{equation}
    A_0=0,\qquad A_L=0.
\end{equation}
The effective energy density can then be decomposed as
\begin{equation}
    \rho_{T}
    =
    \rho_{\mathrm{compact}}^{(T)}
    +
    \rho_{\mathrm{non\text{-}minimal}}^{(T)}.
\end{equation}
The compact contribution to the spectral energy density of the transverse mode is then given by
\begin{equation}
    \rho_{\mathrm{compact}}^{(T)}
    =
    2\left[
    |A'_T|^2+D|A_T|^2
    \right].
\end{equation}
where 
\begin{equation}
	D = k^2 + M^2 a^2
\end{equation}
The non-minimal contribution is given by
\begin{equation}
    \rho_{\mathrm{non\text{-}minimal}}^{(T)}
    =
    2\xi
    \left[
    12\mathcal H\,\mathrm{Re}\!\left(A'_T A_T^*\right)
    -
    F \, |A_T|^2
    \right],
\end{equation}
where
\begin{equation}
	F = a^2R+6\mathcal H^2
	=
	\begin{aligned}
		6(2+\gamma)\mathcal H^2.
	\end{aligned}
	\label{eq:F_definition}
\end{equation}
The parameter \(\gamma\) in Eq.~\eqref{eq:F_definition} characterizes the background evolution and allows the relation to describe the dS,qdS, and RD epochs, according to the definition given in Eq.~\eqref{eq:gamma_definition}. To express the energy density in terms of spectral quantities, we define the cross-spectrum as follows:
\begin{equation}
    \mathcal{C}_{X,Y}(k)
    \equiv
    \frac{k^3}{2\pi^2}
    \mathrm{Re}\!\left[X(k)Y^*(k)\right].
    \label{eq:cross_spectrum}
\end{equation}
The power spectra are defined as
\begin{equation}
    \mathcal{P}_{A}(k)
    \equiv \frac{k^3}{2\pi^2}|A(k)|^2,
    \qquad
    \mathcal{P}_{A'}(k)
    \equiv \frac{k^3}{2\pi^2}|A'(k)|^2.
    \label{eq:power_spectra}
\end{equation}
These spectral quantities allow us to express the energy density as a contribution per logarithmic wavenumber interval, facilitating the analysis of the scale dependence and evolution of the different contributions. Using these definitions, the energy density can be written as
\begin{equation}
    \rho_{T}
    =
    \frac{1}{a^4}
    \int d\ln k
    \left[
    \mathcal{P}_{A'_T}
    +
    D\mathcal{P}_{A_T}
    +
    \xi
    \left(
    12\mathcal H\mathcal{C}_T
    -
    (a^2R+6\mathcal H^2)\mathcal{P}_{A_T}
    \right)
    \right].
    \label{eq:rho_eff_T_spectral}
\end{equation}

\subsubsection*{Longitudinal Mode}
Using the constraint equation \eqref{eq:A0_constraint} to eliminate the temporal component $A_0$, the expression is reduced as follows:
\begin{equation}
    \begin{aligned}
        \rho_{L}(\mathbf{k})
        ={}&
        \frac{M^2a^2}{D}|A'_L|^2
        +
        M^2a^2|A_L|^2
        \\
        &+
        \xi\Bigg\{
        12\mathcal H
        \left(1+\frac{k^2}{D}\right)
        \mathrm{Re}\!\left(A'_L A_L^*\right)
        +
        \frac{24\mathcal H k^2S}{D^2}|A'_L|^2
        \\
        &\qquad+
        \left(a^2R+6\mathcal H^2\right)
        \left[
        \frac{k^2}{D^2}|A'_L|^2-|A_L|^2
        \right]
        \Bigg\}.
    \end{aligned}
    \label{eq:rho_eff_L_A0_eliminated}
\end{equation}
where we define \(S\) as
\begin{equation}
    S
    =
    \frac{M'}{M}
    +
    \mathcal H.
    \label{eq:S_definition}
\end{equation}
Using the spectral quantities defined above, the spectral energy density of the transverse mode is given by

\begin{equation}
	\frac{d\left\langle\rho_{T}\right\rangle}
	{d\ln k}
	=
	\frac{1}{a^4}
		\left[
			\mathcal{P}_{A'_T}
			+
			D\mathcal{P}_{A_T}
			+
			\xi
			\left(
			12\mathcal H\mathcal{C}_T
			-
			(a^2R+6\mathcal H^2)\mathcal{P}_{A_T}
			\right)
			\right].
	\label{eq:spectral_rho_T_compact}
\end{equation}
\noindent
Similarly, the spectral energy density of the longitudinal mode
can be written as
\begin{equation}
    \begin{aligned}
        \frac{d\left\langle\rho_L\right\rangle}
        {d\ln k}
        =
        \frac{1}{2a^4}
        \Bigg\{&
        \frac{M^2a^2}{D}\mathcal{P}_{A'_L}
        +
        M^2a^2\mathcal{P}_{A_L}
        \\
        &+
        \xi\Bigg[
        12\mathcal H
        \left(1+\frac{k^2}{D}\right)\mathcal C_L
        +
        \frac{24\mathcal H k^2S}{D^2}
        \mathcal P_{A'_L}
        \\
        &\qquad+
        \left(a^2R+6\mathcal H^2\right)
        \left(
        \frac{k^2}{D^2}\mathcal P_{A'_L}
        -\mathcal P_{A_L}
        \right)
        \Bigg]
        \Bigg\}.
    \end{aligned}
    \label{eq:spectral_rho_L_compact}
\end{equation}

\noindent
The expressions in 
\eqref{eq:spectral_rho_T_compact} and
\eqref{eq:spectral_rho_L_compact} represent the bare spectral
energy contributions derived from the Hilbert stress-energy tensor.

\noindent
It is important to distinguish these bare energy spectra from a
particle-number spectrum. In particular, the late-time number spectrum is a different observable, given by
\begin{equation}
N(\tilde{k})
=
\frac{\tilde{k}^3}{2\pi^2}
\left(2n_T+n_L\right),
\label{eq:particle_number_spectrum}
\end{equation}
where $n_T$ and $n_L$ are the occupation numbers of one transverse
polarization and the longitudinal polarization, respectively. It cannot be used as a direct replacement for the
spectral energy contributions derived from the Hilbert stress tensor.

\noindent
For completeness, the non-minimal contribution to the Hilbert
stress-energy tensor, in our conventions, can be written  \cite{GrzadkowskiSocha2025,Koivisto2009} 
\begin{equation}
\begin{aligned}
T_{\mu\nu}
={}&T_{\mu\nu}^{\mathrm{Proca}}(m)
+\xi\Big[
G_{\mu\nu}X
+RA_\mu A_\nu
+\left(g_{\mu\nu}\Box
-\nabla_\mu\nabla_\nu\right)X
\Big],
\\
&\qquad X=A_\alpha A^\alpha.
\end{aligned}
\label{eq:nonminimal_stress_tensor}
\end{equation}
Although the Ricci scalar vanishes during radiation domination, the terms
involving $G_{\mu\nu}$ and derivatives of $X$ need not vanish. Thus, the
RD mode equations have $M=m$, while the corresponding
Hilbert energy spectra can retain nonminimal contributions for
$\xi\neq0$.

\noindent
The spectral energy densities considered here are bare quantities and
therefore require an appropriate renormalization prescription before
being interpreted as renormalized physical energy densities.

\noindent
For the constant-$\epsilon$ qdS background, the background
quantities satisfy
\begin{equation}
    a^2R+6\mathcal H^2
    =
    6H_{\rm end}^2(2+\gamma)e^{2\gamma N},
    \qquad
    \gamma=1-\epsilon.
    \label{eq:background_curvature_combination}
\end{equation}

\noindent
The spectral energy densities defined in
Eqs.~\eqref{eq:spectral_rho_T_compact} and
\eqref{eq:spectral_rho_L_compact} are bare, unrenormalized
mode-by-mode contributions obtained from the vacuum expectation
value of the Hilbert stress tensor. Negative values of individual
spectral contributions should therefore not be interpreted as
negative final physical energy densities or, by themselves,
as evidence of ghost instabilities. A physically meaningful
renormalized energy density requires a consistent
renormalization prescription, which is beyond the scope of
the present work.

\subsubsection{dS Case}

For the dS background, the effective mass is constant and
the relevant background quantities satisfy
\begin{equation}
    M'=0,
    \qquad
    S=\mathcal H.
\end{equation}

\subsubsection{qdS Case}

\noindent
For the qdS background, the quantity $S$ is given by
\begin{equation}
S
=
\frac{M'}{M}
+
\mathcal{H}
=
\frac{
\tilde{m}^{2}
+
6\xi\alpha(1+\alpha)a^{-2\epsilon}
}{
\tilde{m}^{2}
+
6\xi(1+\alpha)a^{-2\epsilon}
}
\mathcal{H}.
\end{equation}

\subsubsection{RD Case}

In the RD era, the Ricci scalar and the effective mass reduce to
\begin{equation}
	R=0,
	\qquad
	M^2=m^2,
	\qquad
	S=\mathcal H.
\end{equation}
Because the inflation-to-RD transition is modeled as instantaneous, special care is required when matching the longitudinal mode. The longitudinal reduced action may be written in the form
\begin{equation}
	S_L
	=
	\frac{1}{2}
	\int d\tau\,d^3k
	\left[
	K_L(\tau)|A_L'|^2
	-
	M^2a^2|A_L|^2
	\right].
\end{equation}
Its equation of motion can therefore be written as
\begin{equation}
	\left(K_L A_L'\right)'
	+
	M^2a^2 A_L
	=
	0.
\end{equation}
Integrating this equation across an infinitesimal interval containing the instantaneous transition shows that the canonical momentum, rather than $A_L'$ itself, must be continuous. The appropriate matching conditions are
\begin{equation}
    [A_L]_-^+=0,
    \qquad
    [K_L A_L']_-^+=0.
    \label{eq:longitudinal_matching_conditions}
\end{equation}
Hence, if $M$ changes discontinuously at the end of inflation, the derivative $A_L'$ is in general discontinuous and satisfies
\begin{equation}
	A_L^{\prime +}
	=
	\frac{K_L^-}{K_L^+}
	A_L^{\prime -}.
\end{equation}
These matching conditions should be imposed in the numerical evolution of the longitudinal sector. In a realistic smooth reheating transition, $M_{\mathrm{eff}}$ and $K_L$ vary continuously, and this apparent jump is replaced by a rapid but smooth evolution.

\noindent
To avoid ghost instabilities, the coefficient of $\left|A_L'\right|^2$ must be positive. The standard Maxwell theory is recovered only in the massless and minimally coupled limit, $m=0$ and $\xi=0$, in which only the two transverse degrees of freedom propagate. An accidental cancellation satisfying $M^2=m^2+\xi R=0$ does not restore the $U(1)$ gauge symmetry; instead, it renders the reduced longitudinal action singular.

\noindent
The condition $M^2>0$ ensures that the longitudinal kinetic coefficient $K_L$ is positive, thereby excluding a quadratic ghost instability in the reduced longitudinal sector. This condition does not, however, guarantee the positivity of the effective frequency squared or the absence of tachyonic amplification.

\noindent
The details of the calculation are provided in Appendix~\ref{app:A}.

\subsection[Minimally Curvature-Coupled Case]
{Minimally Coupled Case: $\xi=0$}

In the minimally curvature-coupled case, $\xi=0$, the mass reduces to the bare mass, $M=m$. Consequently, the action in Eq.~\eqref{eq:nonminimal_vector_action} reduces to
\begin{equation}
S=\int d^4x\sqrt{-g}\left[-\frac{1}{4}F_{\mu\nu}F^{\mu\nu}-\frac{1}{2}m^2A_\mu A^\mu\right].
\end{equation}
Thus, in the minimally curvature-coupled case ($\xi=0$), the equations of motion given in Eqs.~\eqref{eq:ATT_eom} and \eqref{eq:ALL_eom} reduce to
\begin{equation}
A_{T,i}''+\left(k^2+m^2a^2\right)A_{T,i}=0.
\label{eq:minimal_transverse_eom}
\end{equation}

\begin{equation}
A_L''+2\frac{k^2}{k^2+m^2a^2}\frac{a'}{a}A_L'
+\left(k^2+m^2a^2\right)A_L=0.
\label{eq:minimal_longitudinal_eom}
\end{equation}
For the spectral energy density, the energy--momentum tensor $T_{\mu\nu}$ reduces to the Proca energy--momentum tensor $T_{\mu\nu}^{\mathrm{Proca}}$. Consequently, the transverse contribution to the spectral energy density becomes
\begin{equation}
    a^3 \frac{d\rho_T}{d\ln k} 
    = \frac{k^3}{2\pi^2 a}\left[ 
        \left|A_T'\right|^2 
        + \left(k^2 + m^2 a^2\right)\left|A_T\right|^2 
    \right],
    \label{eq:spectral_energy_transverse}
\end{equation}
and the longitudinal contribution becomes
\begin{equation}
    a^3 \frac{d\rho_L}{d\ln k} 
    = \frac{k^3}{4\pi^2 a}\left[ 
        \frac{m^2 a^2}{k^2 + m^2 a^2}\left|A_L'\right|^2 
        + m^2 a^2 \left|A_L\right|^2 
    \right].
    \label{eq:spectral_energy_longitudinal}
\end{equation}

\subsection{Ghost and Tachyonic Stability Conditions}
We discuss the stability conditions of the longitudinal mode \cite{Himmetoglu2009a,Himmetoglu2009Ghost,EspositoFarese2010}, distinguishing between ghost and tachyonic instabilities. In the parameter range considered here, the effective mass squared remains positive, $M^2>0$. However, this condition alone does not guarantee the dynamical stability of the longitudinal mode. The squared effective frequency can become negative, $\omega_L^2<0$, leading to a transient tachyonic regime. Such a regime can produce non-oscillatory evolution and amplification of the longitudinal mode, but it does not, by itself, indicate a ghost instability. The absence of a quadratic ghost must instead be assessed from the sign of the longitudinal kinetic coefficient.

\subsubsection{Minimal coupling case}

Using Eq.~\eqref{eq:minimal_longitudinal_eom}, the squared effective frequency of the longitudinal mode can be written, for the three background regimes dS, qdS, and RD, as

\begin{equation}
\omega_L^2 =
k^{2} + m^{2} e^{2N}
- \frac{k^{2} (1 + \gamma) H_{\rm end}^2}{k^{2} + m^{2} e^{2N}}
\, e^{2 \gamma N}
+ \frac{3 m^{2} k^{2} H_{\rm end}^2}{\left(k^{2} + m^{2} e^{2N}\right)^{2}}
e^{2 (1 + \gamma) N}
\label{eq:omega_L2}
\end{equation}

\noindent
where $\gamma$ is defined in Eq.~\eqref{eq:gamma_definition} and covers three scenarios: the dS, qdS, and RD epochs.
In the limit $k \to \infty$, Eq.~\eqref{eq:omega_L2} is dominated by the $k^2$ term, whose coefficient is unity; therefore, there is no gradient instability. In the small-momentum limit, $k^2 \ll m^2 a^2$, the positive mass term, $m^2 a^2$, dominates. Therefore, negative values of $\omega_L^2$ can occur only in an intermediate regime, where the curvature-induced term competes with the positive gradient and mass terms, indicating a tachyonic instability.

\noindent
In this work, we focus on the dS and qdS regimes, computing and comparing the squared effective frequency of the longitudinal mode to assess deviations between them and identifying possible tachyonic or gradient instabilities.

\subsubsection{Non-minimal coupling case}
Using Eq.~\eqref{eq:ALL_eom}, the squared effective frequency of the longitudinal mode can be expressed in the dS, qdS, and RD regimes in terms of $S$, defined in Eq.~\eqref{eq:S_definition}. Its derivative with respect to $N$ is given by

\begin{equation}
S'(N(\tau))
=
\frac{M''}{M}
+
\frac{a''}{a}
-
\left(
\frac{M'}{M}
\right)^{2}
-
\left(
\frac{a'}{a}
\right)^{2}.
\end{equation}

\noindent
Thus, the dimensionless effective frequency of the longitudinal mode acquires additional contributions arising from the time dependence of $\tilde{M}_{\rm eff}$.
\begin{equation}
\begin{aligned}
\omega_L^2(N(\tau)) ={}&
\frac{-k^{4}
+2k^{2}M^{2}a^{2}}
{
\left(k^{2}+M^{2}a^{2}\right)^{2}}
S^2(N(\tau))
\\
&-
\frac{k^{2}}
{
\left(k^{2}+M^{2}a^{2}\right)}
S'(N(\tau)) +
\left(
k^{2}
+M^{2}a^{2}
\right).
\end{aligned}
\label{eq:omega_L2_eff}
\end{equation}
The sign of $\omega_L^2$ provides a diagnostic of the instantaneous oscillatory or tachyonic character of the longitudinal mode, whereas the ghost condition is determined independently by the kinetic coefficient $K_L$. Therefore, the occurrence of $\omega_L^2<0$ does not necessarily signal a ghost instability, provided that the kinetic coefficient remains positive. In the following analysis, these two stability criteria are examined separately.
\noindent
It should be noted that, in the minimally coupled case ($\xi=0$), 
Eq.~\eqref{eq:omega_L2_eff} reduces to Eq.~\eqref{eq:omega_L2}. 
To compare and plot the two effective frequencies, we normalize them by 
$H_{\rm end}^2$, defining
\begin{equation}
    \tilde{\omega}_L^2
    \equiv
    \frac{\omega_L^2}{H_{\rm end}^2}.
    \label{eq:omega_L2_tilde}
\end{equation}

\section{Numerical Computation}

In this section, we numerically compute and analyze the deviation of the background energy density from dS case, as well as the evolution of the transverse and longitudinal modes.

\subsection{Background Energy-Density Deviation}

\begin{figure}[H]
\centering
\includegraphics[width=1.0\textwidth]{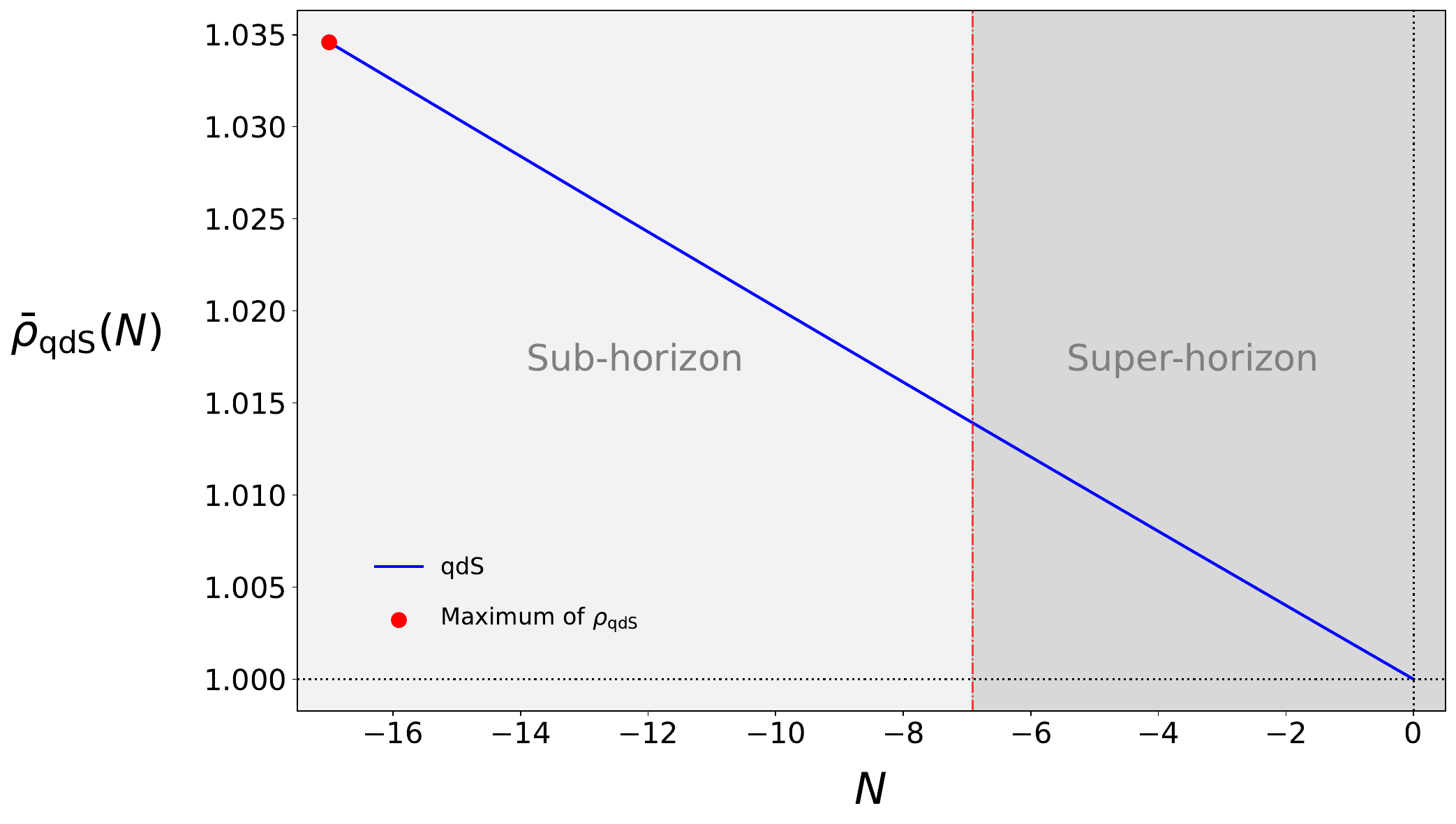}
\caption{
Normalized qdS background energy density,
$\bar{\rho}_{\rm qdS}(N)$, as a function of the e-fold number
$N$ for modes with $k = 10^{-3}H$ and $m = 10^{-2}H$. The horizontal dotted line represents the normalized dS
reference value,
$\bar{\rho}_{\rm dS}=\rho_{\rm dS}/(3M_{\rm Pl}^{2}H_{\rm end}^{2})=1$.
The qdS background energy density is evaluated using
Eq.~\eqref{eq:energy_density_qdS_RD}, while the dS reference
is obtained from Eq.~\eqref{eq:energy_density_dS_RD}.
The vertical dash-dotted line indicates qdS horizon crossing,
and the vertical dotted line marks the end of inflation at $N=0$.
The shaded regions distinguish the sub-horizon and super-horizon
regimes. The qdS background energy density decreases monotonically
from its maximum value,
$\bar{\rho}_{\rm qdS}^{\rm max}\simeq 1.03458$ at $N=-17$,
to its minimum value,
$\bar{\rho}_{\rm qdS}^{\rm min}=1$ at $N=0$.
}
\label{fig:background_density_deviation}
\end{figure}

In this subsection, we evaluate the normalized qdS background energy density relative to the dS reference during inflation. The qdS background is compared with the corresponding dS result over the interval $-17\leq N\leq0$, using Eqs.~\eqref{eq:energy_density_dS_RD} and
\eqref{eq:energy_density_qdS_RD}. We define the normalized qdS background
energy density as
\begin{equation}
\bar{\rho}_{\rm qdS}(N)
\equiv
\frac{\rho_{\rm qdS}(N)}
{3M_{\rm Pl}^{2}H_{\rm end}^{2}}.
\label{eq:rho_bar_qdS}
\end{equation}
Here, $\bar{\rho}_{\rm qdS}(N)$ denotes the dimensionless qdS background energy
density normalized with respect to the reference scale
$3M_{\rm Pl}^{2}H_{\rm end}^{2}$. The resulting evolution is presented in
Fig.~\ref{fig:background_density_deviation}.

\subsection{Investigation of the Transverse and Longitudinal Energy Densities in dS, qdS, and RD}

In this subsection, we numerically solve the equations of motion for the transverse and longitudinal modes in both dS and qdS backgrounds. The resulting mode functions are used to compute the spectral energy densities of the two polarization sectors. We investigate their time evolution over the interval $-17\leq N\leq 9$, covering the inflationary and RD eras.

\noindent
For the instantaneous-transition prescription, the inflationary and RD solutions are treated separately and connected through the appropriate matching conditions at $N=0$. For the transverse and longitudinal modes, these conditions are
\begin{equation}
    [A_T]_-^+=0,
    \qquad
    [A_T']_-^+=0,
    \label{eq:matching_transverse}
\end{equation}
and \eqref{eq:longitudinal_matching_conditions}
respectively, where
\begin{equation}
    [f]_-^+ \equiv f(0^+)-f(0^-),
\end{equation}
These conditions express the continuity of the transverse mode and its first conformal-time derivative, and of the longitudinal mode and its canonical momentum.\\

\noindent
In addition, we examine the wavenumber dependence of the spectral
energy densities at the transition time, $N=0$. For this purpose,
the spectral energy densities are evaluated over the dimensionless
wavenumber range
\begin{equation}
    10^{-7}\leq \tilde{k}\leq 10^{-3},
\end{equation}
to investigate the scale dependence of the transverse and
longitudinal contributions. The initial conditions are imposed
at sufficiently early times, with the initial time chosen
separately for each wavenumber to ensure that the mode is
well inside the horizon.

\subsubsection{Time Evolution of the Transverse and Longitudinal Spectral Energy Densities for a Fixed Wavenumber}

In this section, we numerically compute and compare the spectral energy
densities of the transverse and longitudinal modes for a fixed
dimensionless comoving wavenumber $\tilde{k}=10^{-3}$, corresponding
to the physical wavenumber $k=10^{-3}H_{\rm end}$. Throughout the
analysis, we fix the dimensionless mass parameter to
$\tilde{m}=10^{-2}$ and consider five values of the non-minimal
coupling parameter,
$\xi=0,10^{-9},10^{-8},10^{-7},10^{-6}$.
With the normalization used here, the representative mode
$\tilde{k}=10^{-3}$ exits the horizon at approximately
$N\simeq\ln\tilde{k}\simeq-6.9$ in the exact dS case. It therefore
corresponds to a mode exiting the horizon only a few e-folds before
the end of inflation and should not be identified with a CMB pivot
scale.  CMB-scale modes exit the horizon roughly $50$--$60$ e-folds before the end of inflation, depending on the reheating history~\cite{Leach2003,Liddle2003},
and would therefore require much smaller values of $\tilde{k}$; observational
constraints on the inflationary parameters are given
in~\cite{Aghanim2020}.

\noindent
Thus, we solve the equations of motion for both the transverse and longitudinal modes in the dS and qdS backgrounds, adopting $\epsilon=10^{-3}$.

\noindent
Using Eqs.~\eqref{eq:spectral_rho_T_compact} and \eqref{eq:spectral_rho_L_compact}, we evaluate the spectral energy densities of the transverse and longitudinal modes in both the dS and qdS backgrounds.

\noindent
In addition, to quantify the effect of the slow-roll correction on the spectral energy density for each value of the non-minimal coupling parameter $\xi$, we define the exact difference between the dS and qdS results as

\begin{equation}
    \Delta X \equiv X_{\rm qdS}-X_{\rm dS}.
    \label{eq:difference}
\end{equation}
This quantity provides a direct measure of the slow-roll correction to the spectral energy density and allows us to examine how the deviation between the dS and qdS results varies with the non-minimal coupling parameter $\xi$ during inflation. We subsequently plot this quantity as a function of the e-fold number $N$ over the considered range of coupling strengths $\xi$.

\noindent
Furthermore, for each value of the non-minimal coupling parameter $\xi$, we determine the e-fold number at which the effective mass satisfies the mass-crossing condition,
\begin{equation}
	\tilde{M}(N_{\rm root})
	=
	\frac{\tilde{k}}{a(N_{\rm root})},
	\label{eq:mass_crossing}
\end{equation}
\noindent
Here, the mass-crossing time is defined as the e-fold time at which
the effective mass equals the physical wavenumber, $M=k/a$.
We denote this time by $N_{M=k/a}$. The corresponding spectral energy
densities are evaluated at their respective mass-crossing times,
$N_{M=k/a}^{\rm dS}$ and $N_{M=k/a}^{\rm qdS}$. Thus, the comparison
is performed using the same physical mass-crossing criterion, $M=k/a$,
rather than at the same e-fold time. In the following, we define the
dimensionless quantities by
\begin{equation}
\tilde{\rho} \equiv \frac{\rho}{H_{\rm end}^{4}},
\qquad
\tilde{m} \equiv \frac{m}{H_{\rm end}},
\qquad
\tilde{k} \equiv \frac{k}{H_{\rm end}},
\qquad
\tilde{M} \equiv \frac{M}{H_{\rm end}}.
\end{equation}

\subsubsection{Transverse Mode}
\noindent
In this section, we investigate the spectral energy density of the transverse mode given by Eq.~\eqref{eq:spectral_rho_T_compact}, for various values of the non-minimal coupling parameter $\xi$ in both dS and qdS backgrounds. To quantify the effects of slow-roll corrections, we compute the exact difference, $\Delta$, between the dS and qdS spectral energy densities, defined in Eq.~\eqref{eq:difference}.

\begin{figure}[H]
\centering
\includegraphics[width=1.0\textwidth]{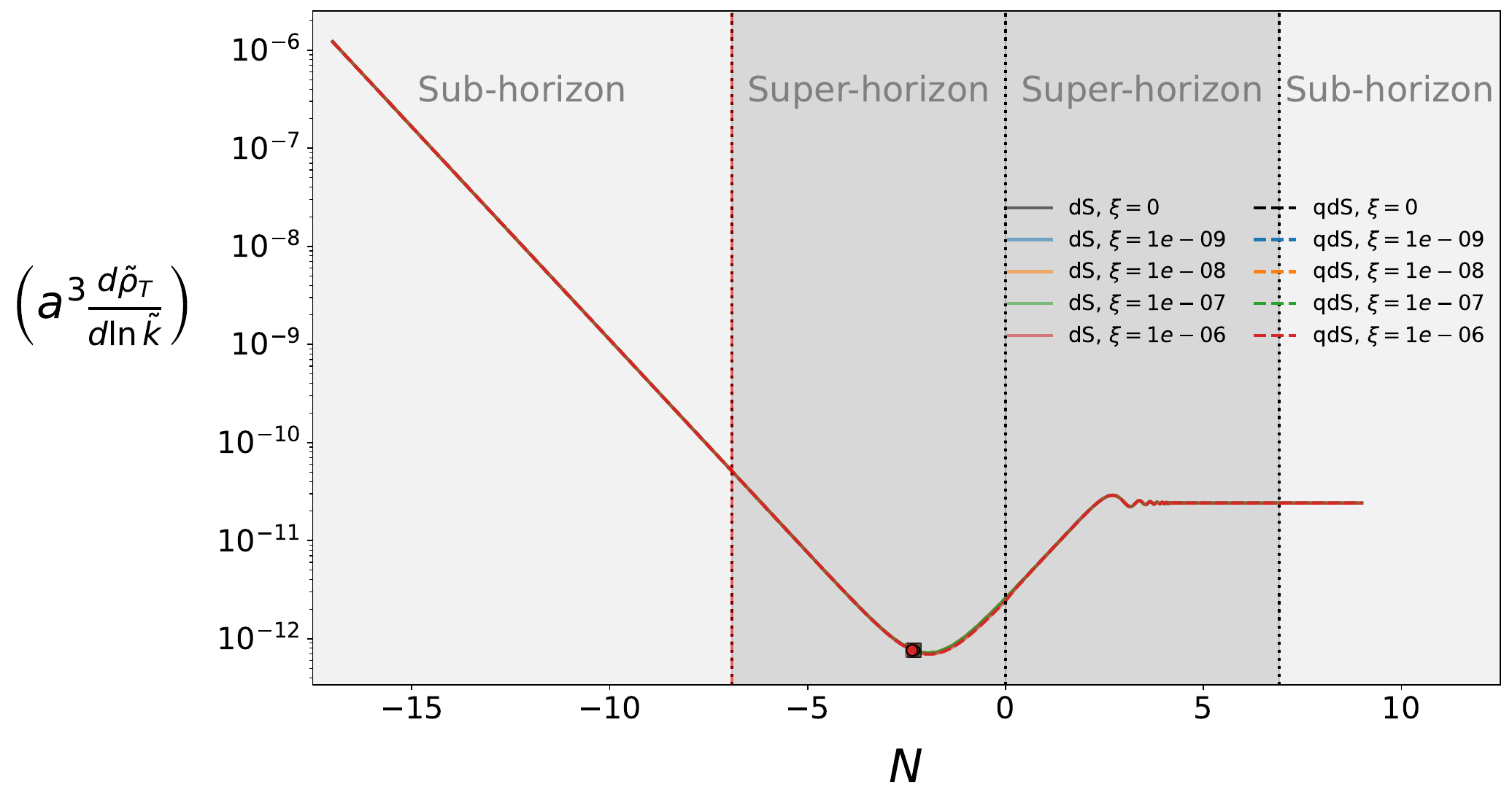}
\caption{
Time evolution of the transverse-mode spectral energy density,
$a^3\,d\tilde{\rho}_T/d\ln\tilde{k}$, for
$\xi=0,10^{-9},10^{-8},10^{-7},10^{-6}$ and
$\epsilon=10^{-3}$.
The effective mass is given by
$\tilde{M}^{2}=\tilde{m}^{2}+12\xi$ in the dS regime and
$\tilde{M}^{2}=\tilde{m}^{2}+6\xi(1+\alpha)e^{-2\epsilon N}$
in the qdS regime.
The vertical lines indicate the characteristic times during the mode
evolution: horizon crossing in the dS regime at
$N=\ln(\tilde{k})=-6.90776$, horizon crossing in the qdS regime at
$N=\ln(\tilde{k})/\alpha=-6.91467$, the transition from the
inflationary phase to the RD epoch at $N=0$,
and horizon re-entry during RD at
$N=\ln(1/\tilde{k})=6.90776$.
The instantaneous-transition prescription produces a discontinuous
change in the curvature-induced contribution to $M^2$ at $N=0$,
since the Ricci scalar vanishes in the RD epoch.
The inflationary and RD solutions are therefore evaluated and plotted
as separate one-sided branches, without introducing an artificial
interpolation across the transition. Consequently, any discontinuity
in the spectral energy density at $N=0$ is retained.
A smooth reheating transition would instead produce a continuous
evolution of the curvature-induced contribution to the effective mass.
Throughout the considered coupling values, the transverse-mode
spectral energy density remains positive in the dS and qdS
inflationary branches.
}
\label{fig:rho_T_Total}
\end{figure}

\noindent
Figure~\ref{fig:rho_T_Total} shows the evolution of the unnormalized spectral energy density of the transverse mode given by Eq.~\eqref{eq:spectral_rho_T_compact} for different values of the non-minimal coupling parameter $\xi$ in both the dS and qdS backgrounds. For the minimally coupled case ($\xi=0$), the mode evolution is continuous across $N=0$ within each dS$\rightarrow$RD and qdS$\rightarrow$RD history. However, as $\xi$ increases, the effective mass in the qdS background acquires a time dependence while remaining constant in the dS case. Consequently, the difference between the dS and qdS unnormalized spectral energy densities near the transition at $N=0$ becomes progressively larger with increasing $\xi$, demonstrating the growing influence of the non-minimal curvature coupling on the evolution of the transverse mode. Throughout the considered coupling range, $0\leq\xi\leq10^{-6}$, the unnormalized transverse-mode spectral energy density remains positive in both the dS and qdS backgrounds.

\begin{figure}[H]
    \centering
    \includegraphics[width=1.0\textwidth]{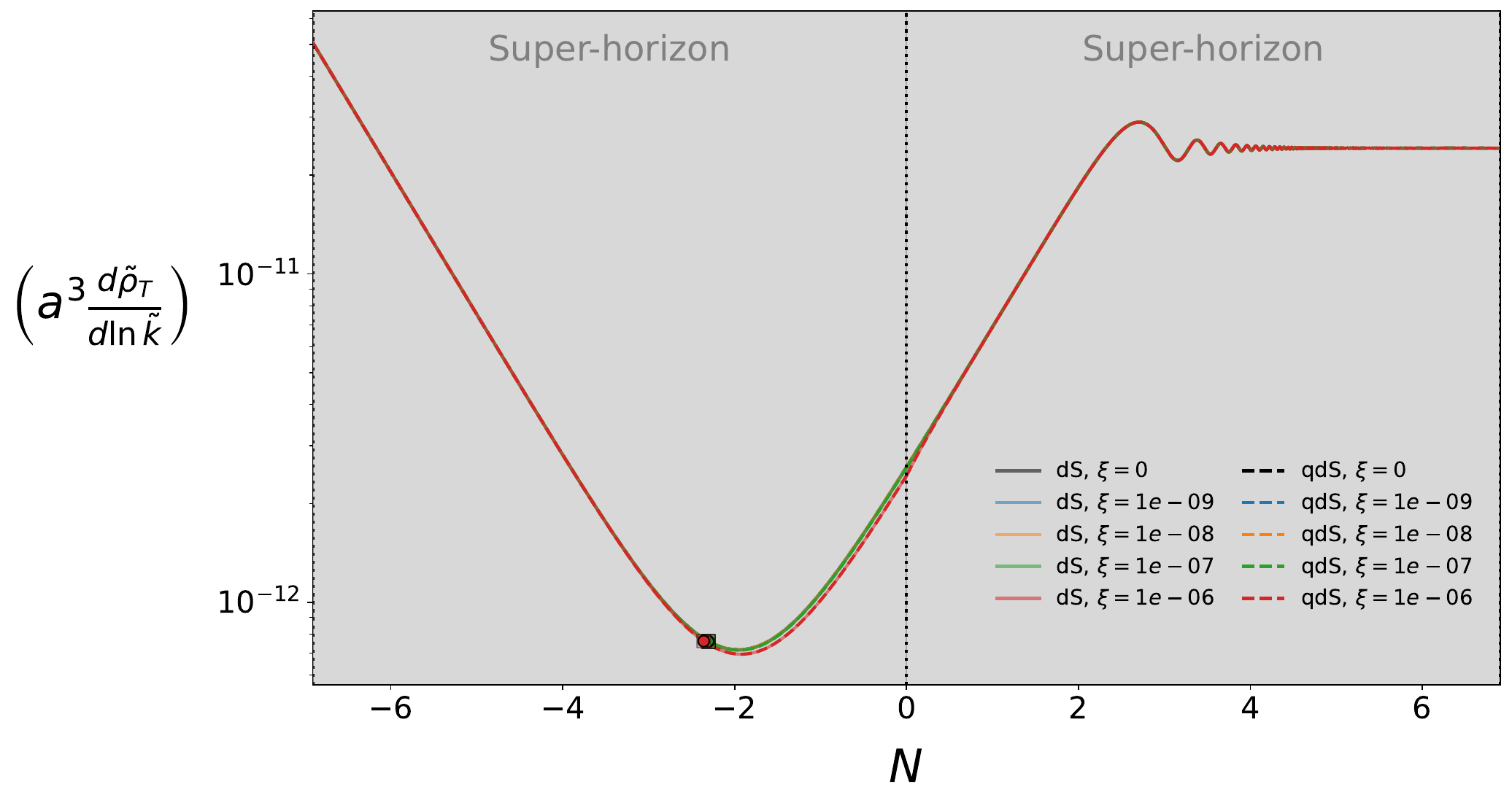}
    \caption{
    Superhorizon evolution of the transverse-mode spectral energy
    density, $a^3\,d\tilde{\rho}_T/d\ln\tilde{k}$, from horizon
    exit during inflation to horizon re-entry in the 
    RD epoch, for $\xi=0,10^{-9},10^{-8},10^{-7},10^{-6}$
    and $\epsilon=10^{-3}$.
    The displayed interval,
    $\ln(\tilde{k})\leq N\leq\ln(1/\tilde{k})$,
    corresponds to the superhorizon regime.
    The solid and dashed curves represent the dS and qdS
    inflationary backgrounds, respectively, together with their
    corresponding RD evolutions.
    The vertical line at $N=0$ marks the instantaneous transition
    from inflation to RD.
    Since the curvature-induced contribution to the effective mass
    vanishes abruptly when the Ricci scalar becomes zero in RD,
    the inflationary and RD solutions are evaluated as separate
    one-sided branches, without interpolation across the transition.
    Consequently, the discontinuity in the spectral energy density
    at $N=0$ is retained.
    }
    \label{fig:rho_T_Total_sup}
\end{figure}

\noindent
Table~\ref{tab:rhoT_masscrossing_small_xi} presents the e-fold
time $N_{M=k/a}$ at which the mass-crossing condition
Eq.~\eqref{eq:mass_crossing} is satisfied, together with
the corresponding transverse-mode spectral energy density
Eq.~\eqref{eq:spectral_rho_T_compact} for different values of the
non-minimal coupling parameter $\xi$ in the dS and qdS backgrounds.

\noindent
The calculations use $\tilde{m}=0.01$, $\tilde{k}=10^{-3}$,
and $a_{\rm end}=1$. The spectral energy density is
normalized by $H_{\rm end}^{4}$, and the quantity
reported in the table is
\begin{equation}
\frac{a^3}{H_{\rm end}^{4}}
\frac{d\rho_{T,\mathrm{bare}}}{d\ln k},
\label{eq:rhoT_masscrossing_normalization}
\end{equation}
including both transverse polarizations and evaluated
at the corresponding mass-crossing time. The
mass-crossing times are determined independently of
the spectral-energy-density evaluation: analytically
in the dS background and numerically, using the
Brent root-finding method~\cite{Brent1971}, in the qdS background.

\noindent
For the initial conditions, we specify the initial
subhorizon through the dimensionless parameters

\begin{equation}
Q_{\mathrm{depth}}^{\mathrm{dS}}
=\tilde{k}e^{17},
\qquad
Q_{\mathrm{depth}}^{\mathrm{qdS}}
=\tilde{k}e^{17\alpha},
\label{eq:Q_depth}
\end{equation}
which determine the initial times as
\begin{equation}
N_{\mathrm{ini}}^{\mathrm{dS}}
=\ln\left(
\frac{\tilde{k}}{Q_{\mathrm{depth}}^{\mathrm{dS}}}
\right),
\qquad
N_{\mathrm{ini}}^{\mathrm{qdS}}
=\frac{1}{\alpha}
\ln\left(
\frac{\tilde{k}}{Q_{\mathrm{depth}}^{\mathrm{qdS}}}
\right).
\label{eq:initial_times_depth}
\end{equation}
With these choices, both backgrounds are initialized
at $N=-17$, corresponding to 17 e-folds before the end
of inflation. Second-order WKB Bunch--Davies initial conditions~\cite{Baumann2009,Riotto2003}
are imposed at these initial times.

\noindent
The results show that increasing $\xi$ shifts the
mass-crossing time to earlier epochs, as indicated by
the increasingly negative values of $N_{M=k/a}$.
Over the same range of couplings, the transverse-mode
spectral energy density increases monotonically in both
backgrounds.

\noindent
This means that the spectral energy densities are evaluated at different mass-crossing times, depending on $\xi$. Furthermore, the qdS values
are slightly larger than their dS counterparts for
every nonzero value of $\xi$ considered. This indicates
a small enhancement of the transverse-mode spectral
energy density due to the slow-roll correction for the
parameters examined.

\begin{table}[H]
\centering
\small
\setlength{\tabcolsep}{3pt}
\renewcommand{\arraystretch}{1.1}

\caption{
The e-fold time $N_{M=k/a}$ at which the effective mass
satisfies $M=k/a$, and the corresponding transverse-mode
spectral energy density,
$a^3 d\tilde{\rho}_T/d\ln\tilde{k}$, for different values
of the non-minimal coupling parameter $\xi$ in dS and qdS
backgrounds. The spectral energy density includes both
transverse polarizations and is evaluated at the respective
mass-crossing time in each background. The calculations
use $\tilde{m}=0.01$, $\tilde{k}=10^{-3}$,
$\epsilon=10^{-3}$, and $a_{\rm end}=1$.
}
\label{tab:rhoT_masscrossing_small_xi}

\begin{tabular}{|c|c|c|c|c|}
\hline
$\xi$
&
$N_{M=k/a}^{\rm dS}$
&
$\begin{array}{c}
\left(a^3 d\tilde{\rho}_T/d\ln\tilde{k}\right)_{\rm dS}
\end{array}$
&
$N_{M=k/a}^{\rm qdS}$
&
$\begin{array}{c}
\left(a^3 d\tilde{\rho}_T/d\ln\tilde{k}\right)_{\rm qdS}
\end{array}$
\\
\hline

$0$
&
$-2.30258509$
&
$7.59762395\times10^{-13}$
&
$-2.30258509$
&
$7.59763497\times10^{-13}$
\\

$1\times10^{-9}$
&
$-2.30264509$
&
$7.59762400\times10^{-13}$
&
$-2.30264534$
&
$7.59763494\times10^{-13}$
\\

$1\times10^{-8}$
&
$-2.30318473$
&
$7.59762574\times10^{-13}$
&
$-2.30318720$
&
$7.59763600\times10^{-13}$
\\

$1\times10^{-7}$
&
$-2.30854938$
&
$7.59776342\times10^{-13}$
&
$-2.30857384$
&
$7.59776786\times10^{-13}$
\\

$1\times10^{-6}$
&
$-2.35924944$
&
$7.60987594\times10^{-13}$
&
$-2.35947587$
&
$7.60991171\times10^{-13}$
\\

\hline
\end{tabular}
\end{table}

\begin{figure}[H]
\centering
\includegraphics[width=1.0\textwidth]{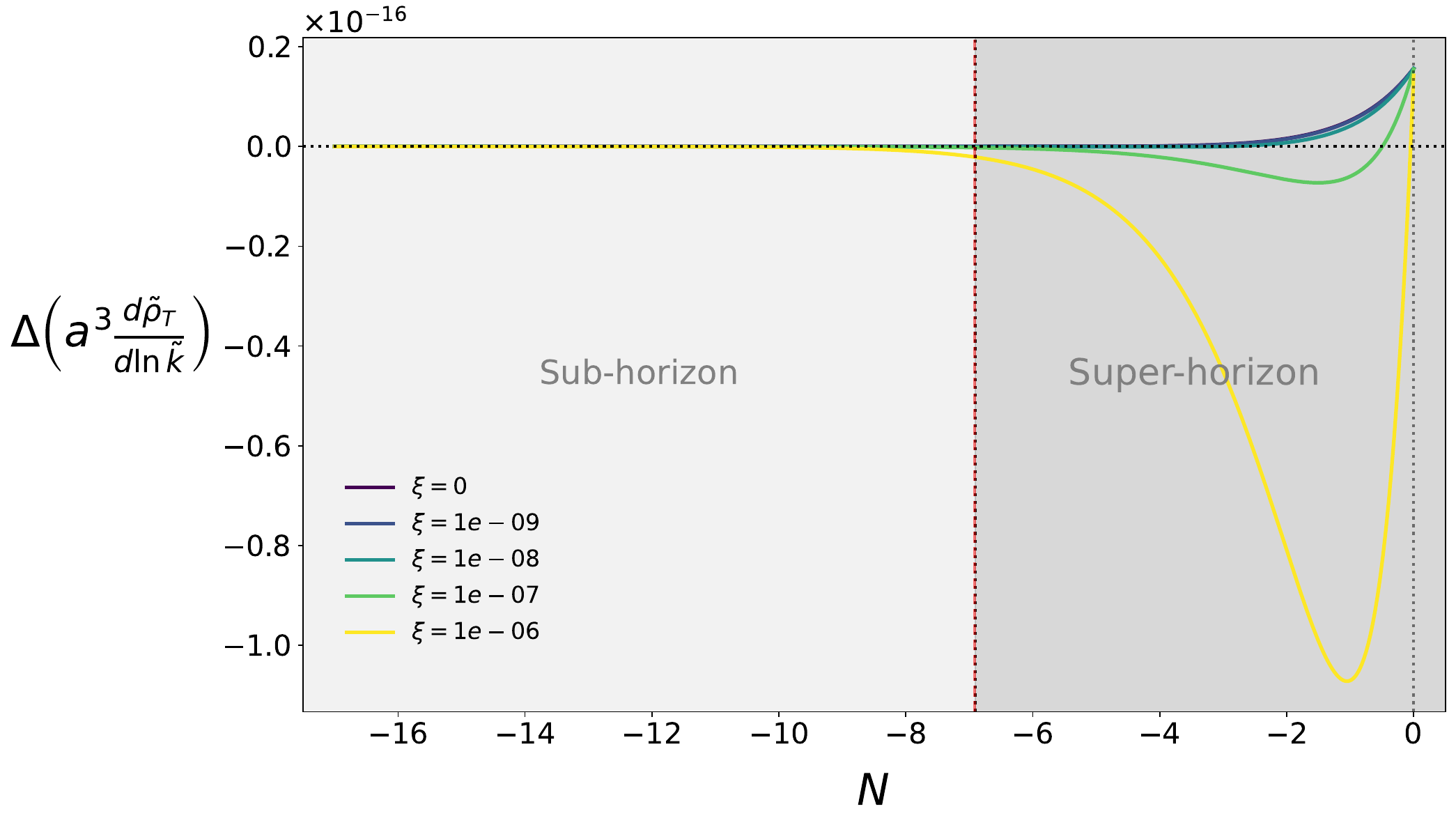}
\caption{
Difference between the transverse-mode spectral energy densities
in the qdS and dS backgrounds, defined as \eqref{eq:difference}
for $0\leq\xi\leq10^{-6}$ and $\epsilon=10^{-3}$.
The effective mass is given by
$\tilde{M}^{2}=\tilde{m}^{2}+12\xi$ in the dS background and
$\tilde{M}^{2}=\tilde{m}^{2}+6\xi(1+\alpha)e^{-2\epsilon N}$
in the qdS background.
The vertical lines indicate the horizon-crossing times in the dS
and qdS backgrounds, at
$\ln(\tilde{k})=-6.90775$ and
$\ln(\tilde{k})/\alpha=-6.91466$, respectively, and the transition
to the RD epoch at $N=0$.
}
\label{fig:difference_rho_T_total}
\end{figure}

\noindent
Figure~\ref{fig:difference_rho_T_total} shows the difference between
the transverse-mode spectral energy densities in the qdS and dS
backgrounds, defined in Eq.~\eqref{eq:difference}, for different
values of the non-minimal coupling parameter $\xi$. The difference
is evaluated over the inflationary period by interpolating the qdS
spectral energy density onto the dS e-fold grid.

\noindent
The details of the calculation are provided in Appendix~\ref{app:B}.

\subsubsection{Longitudinal Mode}

We now turn our attention to the longitudinal mode, whose spectral
energy density is given by Eq.~\eqref{eq:spectral_rho_L_compact}.
We investigate its behavior for various values of the non-minimal
coupling parameter $\xi$ in both dS and qdS backgrounds.

\noindent
Compared to the transverse modes, the longitudinal mode exhibits
a strong sensitivity to the curvature coupling, since both its
kinetic normalization and its effective frequency depend explicitly
on the effective mass $M$. For the positive couplings considered
in this work, the Ricci scalar contribution enhances the effective
mass during inflation. This enhancement modifies the epoch at which
the physical momentum of a given mode becomes comparable to its
effective mass. The spectral energy densities presented below
illustrate the consequences of this modification and allow us to
compare the longitudinal-mode evolution in the dS and qdS regimes.

\begin{figure}[H]
\centering
\includegraphics[width=1.0\textwidth]{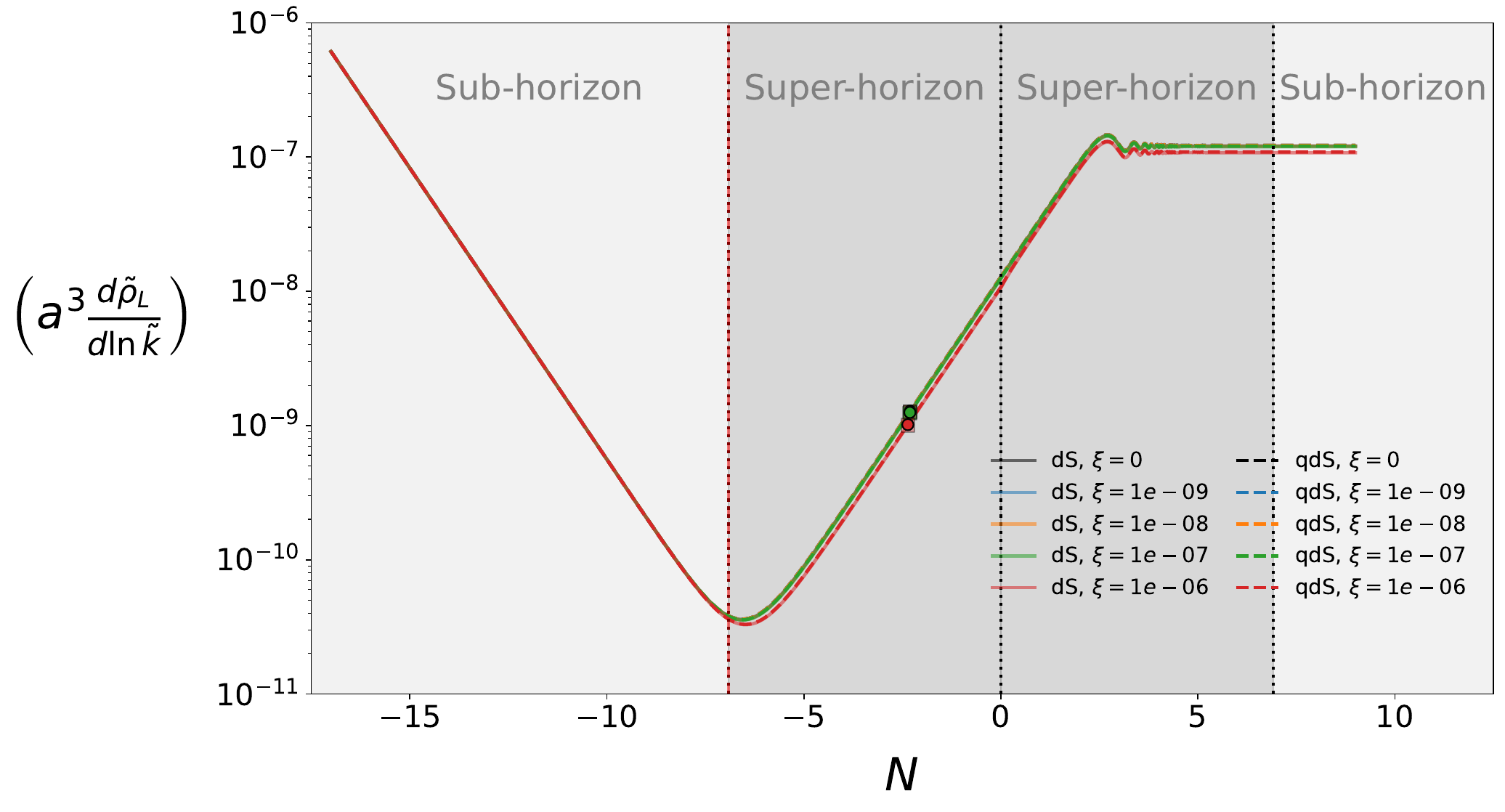}
\caption{
Spectral energy density of the longitudinal mode for the range
$0\leq\xi\leq10^{-6}$ with $\epsilon=10^{-3}$.
The solid and dashed curves correspond to the dS and qdS
backgrounds, respectively. The effective mass is defined as
$\tilde{M}^{2}_{\mathrm{dS}}=\tilde{m}^{2}+12\xi$ in the dS
background and
$\tilde{M}^{2}_{\mathrm{qdS}}(N)=\tilde{m}^{2}
+6\xi(1+\alpha)e^{-2\epsilon N}$ in the qdS background.
The vertical lines indicate the characteristic times during
the mode evolution: horizon crossing in the dS background at
$\ln(\tilde{k})=-6.90775$, horizon crossing in the qdS
background at $\ln(\tilde{k})/\alpha=-6.91466$, the transition
from inflation to the RD epoch at $N=0$, and horizon re-entry
during RD at $\ln(1/\tilde{k})=6.90775$.
The sharp feature near $N=0$ results from the instantaneous
transition approximation. Since the Ricci scalar vanishes
during RD, the curvature-induced contribution to the effective
mass is switched off abruptly at the transition. A smooth
reheating transition would smooth this feature.
}
\label{fig:rho_L_Total}
\end{figure}

\noindent
Figure~\ref{fig:rho_L_Total} shows the evolution of the longitudinal-mode
spectral energy density, given by Eq.~\eqref{eq:spectral_rho_L_compact},
for different values of the non-minimal coupling parameter $\xi$
in both dS and qdS backgrounds. The solid and dashed curves correspond
to the dS and qdS solutions, respectively. For each background,
the inflationary solution is matched to the subsequent RD evolution by imposing \eqref{eq:longitudinal_matching_conditions}.

\noindent
During inflation ($N<0$), the amplitude of the longitudinal-mode
spectral energy density increases with $\xi$ over the range of
couplings considered. In the RD epoch, the dependence on $\xi$
is different, with the plotted amplitude decreasing as $\xi$
increases. This behavior reflects the influence of the curvature
coupling on the inflationary mode evolution and on the initial
conditions inherited by the RD solutions.

\noindent
The dS and qdS backgrounds have different effective-mass evolutions
for nonzero $\xi$. In the dS case, the effective mass is constant,
whereas in the qdS case it depends on time through the slowly
evolving Ricci scalar. Consequently, the two longitudinal-mode
spectra become increasingly distinguishable as $\xi$ increases.
For $\xi=0$, the effective mass reduces to the constant value
$\tilde M=\tilde m$ in both backgrounds, although their different
expansion histories can still lead to differences in the mode
evolution.

\begin{figure}[H]
\centering
\includegraphics[width=1.0\textwidth]{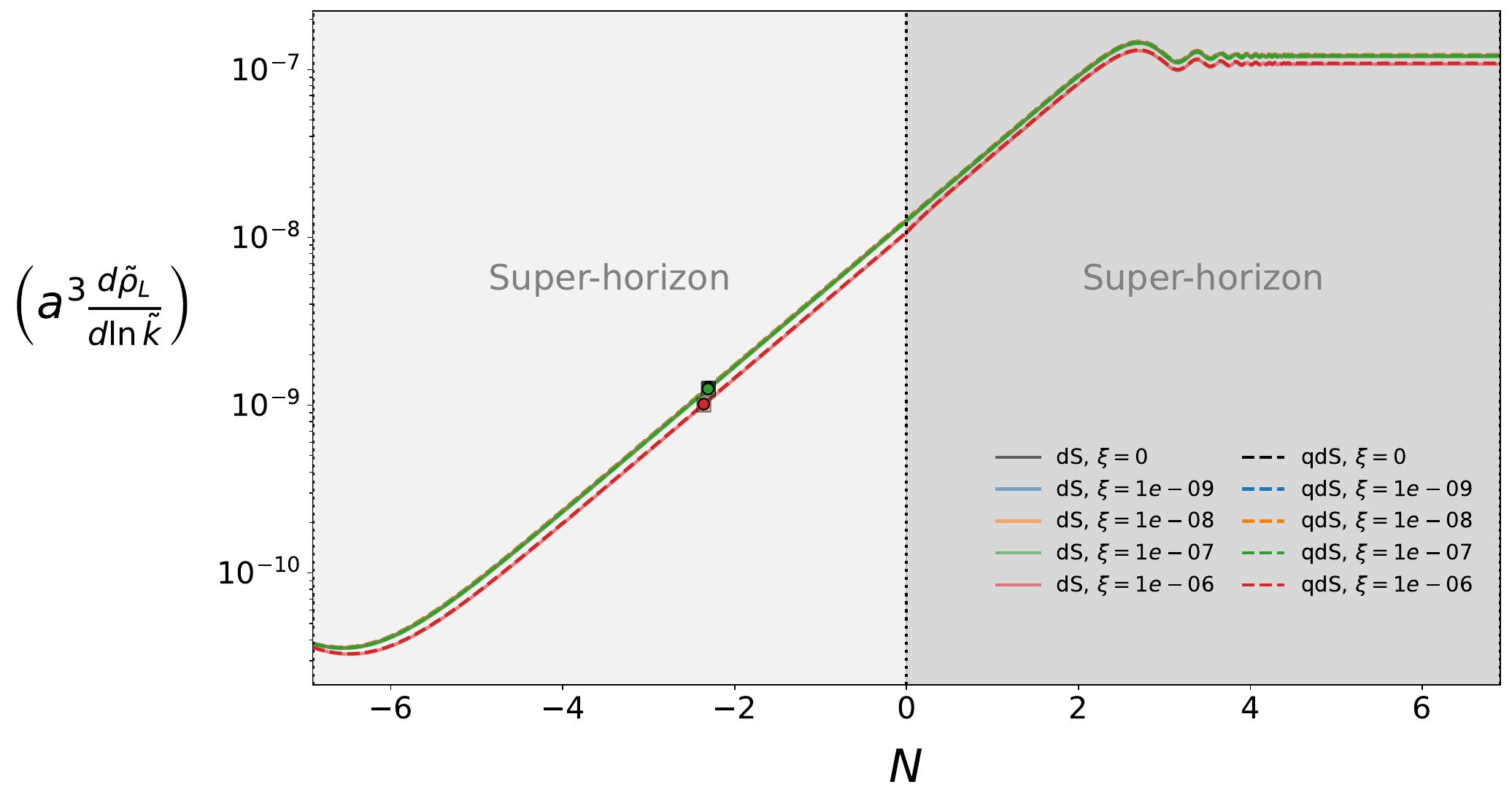}
\caption{
Superhorizon evolution of the longitudinal-mode spectral energy
density from inflation through the transition to RD, for $0\leq\xi\leq10^{-6}$ and
$\epsilon=10^{-3}$. The solid and dashed curves correspond to
the dS and qdS backgrounds, respectively. The effective mass is
given by $\tilde{M}^{2}_{\mathrm{dS}}=\tilde{m}^{2}+12\xi$
in the dS background and
$\tilde{M}^{2}_{\mathrm{qdS}}(N)=\tilde{m}^{2}
+6\xi(1+\alpha)e^{-2\epsilon N}$ in the qdS background.
The vertical line at $N=0$ indicates the transition from
inflation to RD. The sharp feature near the transition results
from the instantaneous transition approximation, in which the
curvature-induced contribution to the effective mass vanishes
abruptly as the Ricci scalar becomes zero during RD.
A smooth reheating transition would smooth this feature.
}
\label{fig:rho_L_Total_sup}
\end{figure}

\noindent
Table~\ref{tab:rhoL_masscrossing} presents the e-fold
time $N_{M=k/a}$ at which the mass-crossing condition
Eq.~\eqref{eq:mass_crossing} is satisfied, together with
the corresponding longitudinal-mode spectral energy
density \eqref{eq:spectral_rho_L_compact} for different values of the non-minimal coupling parameter $\xi$ in the dS and qdS backgrounds. 

\noindent
The calculations are performed for
$\tilde{m}=0.01$, $\tilde{k}=10^{-3}$,
$\epsilon=10^{-3}$, and $a_{\rm end}=1$, considering
the coupling values
$\xi\in\{0,10^{-9},10^{-8},10^{-7},10^{-6}\}$.
The longitudinal sector contains a single polarization.
The spectral energy density \eqref{eq:spectral_rho_L_compact} is evaluated at the corresponding mass-crossing time in each background.

\noindent
The initialization prescription is specified by Eqs.~\eqref{eq:Q_depth} and~\eqref{eq:initial_times_depth}. With these choices, both backgrounds are initialized at $N=-17$, corresponding to 17 e-folds before the end of inflation. Second-order canonical WKB initial conditions are imposed at these initial times.

\noindent
The mass-crossing time is obtained analytically in
the dS background from
\begin{equation}
N_{M=k/a}^{\mathrm{dS}}
=
\ln\left(
\frac{\tilde{k}}
{\sqrt{\tilde{m}^{2}+12\xi}}
\right),
\end{equation}
whereas in the qdS background it is determined
numerically by solving
\begin{equation}
\left[
\tilde{m}^{2}
+6\xi(1+\alpha)e^{-2\epsilon N}
\right]e^{2N}
=\tilde{k}^{2}.
\end{equation}
The corresponding longitudinal-mode spectral energy
densities \eqref{eq:spectral_rho_L_compact} are evaluated from the numerical inflationary
solutions at these crossing times.

\noindent
The mass-crossing times become earlier as the
non-minimal coupling increases in both backgrounds.
The dS and qdS results can be compared to assess
the effect of the slowly varying inflationary
background on the crossing time and on the
longitudinal-mode spectral energy density \eqref{eq:spectral_rho_L_compact}.

\begin{table}[H]
\centering
\footnotesize
\setlength{\tabcolsep}{3pt}
\renewcommand{\arraystretch}{1.05}

\caption{
The e-fold time $N_{M=k/a}$ at which the effective mass
satisfies $M=k/a$, and the corresponding longitudinal-mode
spectral energy density,
$d\tilde{\rho}_{L}/d\ln\tilde{k}$, for different values
of $\xi$ in the dS and qdS backgrounds.
}
\label{tab:rhoL_masscrossing}

\begin{tabular}{|c|c|c|c|c|}
\hline
$\xi$ &
$N_{M=k/a}^{\rm dS}$ &
$a^3\left(d\tilde{\rho}_{L}/d\ln\tilde{k}\right)_{\rm dS}$ &
$N_{M=k/a}^{\rm qdS}$ &
$a^3\left(d\tilde{\rho}_{L}/d\ln\tilde{k}\right)_{\rm qdS}$
\\
\hline

$0$
& $-2.30258509$
& $1.26579687\times10^{-9}$
& $-2.30258509$
& $1.28273565\times10^{-9}$
\\

$1\times10^{-9}$
& $-2.30264509$
& $1.26549310\times10^{-9}$
& $-2.30264534$
& $1.28242486\times10^{-9}$
\\

$1\times10^{-8}$
& $-2.30318473$
& $1.26276355\times10^{-9}$
& $-2.30318720$
& $1.27963219\times10^{-9}$
\\

$1\times10^{-7}$
& $-2.30854938$
& $1.23588783\times10^{-9}$
& $-2.30857384$
& $1.25213952\times10^{-9}$
\\

$1\times10^{-6}$
& $-2.35924944$
& $1.00383545\times10^{-9}$
& $-2.35947587$
& $1.01511067\times10^{-9}$
\\

\hline
\end{tabular}
\end{table}

\begin{figure}[H]
\centering
\includegraphics[width=1.0\textwidth]{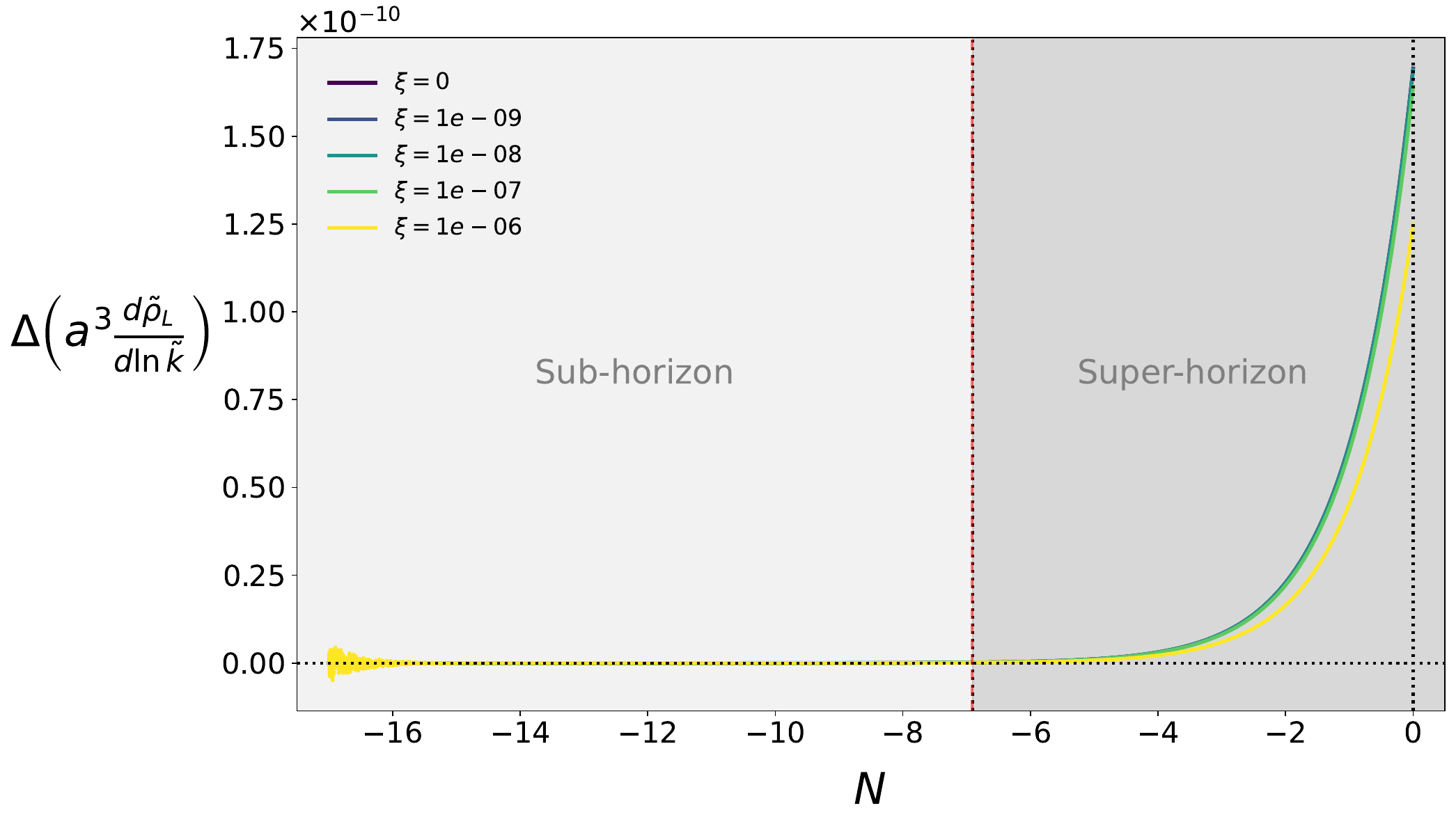}
\caption{
The difference between the longitudinal-mode spectral energy
densities in the qdS and dS backgrounds, defined in
Eq.~\eqref{eq:difference}, for $0 \leq \xi \leq 10^{-6}$ and
$\epsilon = 10^{-3}$. The difference is defined as the qdS
contribution minus the dS contribution. The effective mass is
given by
$\tilde{M}^{2}_{\mathrm{dS}} = \tilde{m}^{2} + 12\xi$
in the dS background and
$\tilde{M}^{2}_{\mathrm{qdS}}(N)
= \tilde{m}^{2} + 6\xi(1+\alpha)e^{-2\epsilon N}$
in the qdS background. The qdS spectral energy density is
interpolated onto the dS e-fold grid to evaluate the difference.
The vertical lines indicate the characteristic times during
the mode evolution: the dS horizon crossing at
$N=\ln(\tilde{k})=-6.90775$, the qdS horizon crossing at
$N=\ln(\tilde{k})/\alpha=-6.91466$, and the transition from
inflation to RD at $N=0$.
}
\label{fig:difference_rho_L_total}
\end{figure}

\noindent
Figure~\ref{fig:difference_rho_L_total} shows the difference between
the longitudinal-mode spectral energy densities in the qdS and dS
spacetimes, defined in Eq.~\eqref{eq:difference}, for different values
of the non-minimal coupling parameter $\xi$. The difference is
defined as the qdS contribution minus the dS contribution and is
evaluated over the inflationary interval by interpolating the qdS
spectral energy density onto the dS e-fold grid.

\noindent
The longitudinal-mode spectral energy densities in the dS and qdS
backgrounds exhibit a time-dependent difference whose magnitude
and sign vary throughout the inflationary evolution. A change in
the sign of this difference indicates a reversal in the relative
ordering of the two spectral contributions, showing that the qdS
correction cannot be described as a simple rescaling of the dS
result. This behavior reflects the combined influence of the
different background evolutions, including the slightly different
horizon-crossing epochs and the time dependence of the
curvature-induced effective mass in the qdS background. The
dependence of the effective mass on $\xi$ further modifies the
evolution of the longitudinal modes.

\subsection{Spectral Energy Density of the Transverse and Longitudinal Modes as a Function of Wavenumber at $N=0$}
\label{sec:rho_vs_k_N0}

In this section, we investigate the wavenumber dependence of the
spectral energy densities of the transverse
\eqref{eq:spectral_rho_T_compact} and longitudinal
\eqref{eq:spectral_rho_L_compact} modes at a fixed time,
chosen at the transition from inflation to the RD regime ($N=0$).
The analysis is performed over the range
$10^{-7}\leq\tilde{k}\leq10^{-3}$, allowing us to examine the
behavior of the spectral energy densities
\eqref{eq:spectral_rho_T_compact} and
\eqref{eq:spectral_rho_L_compact} across different scales.
For each wavenumber $\tilde{k}$, the mode evolution is initialized
at an initial e-fold time determined by the prescription
\eqref{eq:initial_times_depth} with $Q$ given by
Eq.~\eqref{eq:Q_depth}.
Here, $\epsilon=0$ for the dS background and
$\epsilon=10^{-3}$ for the qdS background.
With these choices, both prescriptions give
\begin{equation}
N_{\rm in}^{\rm dS}
=
N_{\rm in}^{\rm qdS}
=
-17,
\label{eq:Nin_fixed}
\end{equation}
independently of $\tilde{k}$. Thus, all modes are initialized
at the same initial e-fold time in each background.
\noindent
The initial conditions are imposed on the canonical variables
\eqref{eq:K_L}, using normalized positive-frequency
second-order WKB initial data,
\begin{equation}
v_\lambda(N_{\rm in})
=
\frac{1}{\sqrt{2W_\lambda(N_{\rm in})}},
\qquad
v_{\lambda,N}(N_{\rm in})
=
\left[
-iW_\lambda e^{-\mu N}
-\frac{W_{\lambda,N}}{2W_\lambda}
\right]_{N_{\rm in}}
v_\lambda(N_{\rm in}),
\label{eq:WKB_initial_conditions}
\end{equation}
where
\begin{equation}
\mu =
\begin{cases}
1, & \text{dS},\\
\alpha, & \text{qdS},\\
1, & \text{RD}.
\end{cases}
\label{eq:mu_definition}
\end{equation}
where $W_\lambda$ denotes the corresponding second-order
adiabatic frequency of each canonical mode. The resulting mode
functions are evolved from $N=-17$ to the common evaluation
time $N=0$, at which the spectral energy densities are
calculated. The dS calculation is also checked against
the corresponding exact mode solutions.

\subsubsection{Transverse Mode}

In this section, we investigate the dependence of the transverse-mode
spectral energy density, given by Eq.~\eqref{eq:spectral_rho_T_compact},
on the non-minimal coupling parameter $\xi$ in both dS and qdS
backgrounds. The analysis is performed over the range of dimensionless
wavenumbers $10^{-7}\leq\tilde{k}\leq10^{-3}$.

\noindent
To quantify the effect of slow-roll corrections, we calculate the
difference between the qdS and dS spectral energy densities,
$\Delta$, as defined in Eq.~\eqref{eq:difference}. This comparison
allows us to examine how the departure from exact de Sitter expansion
modifies the transverse-mode spectrum as a function of $\tilde{k}$
and $\xi$.

\begin{figure}[H]
\centering
\includegraphics[width=1.0\textwidth]{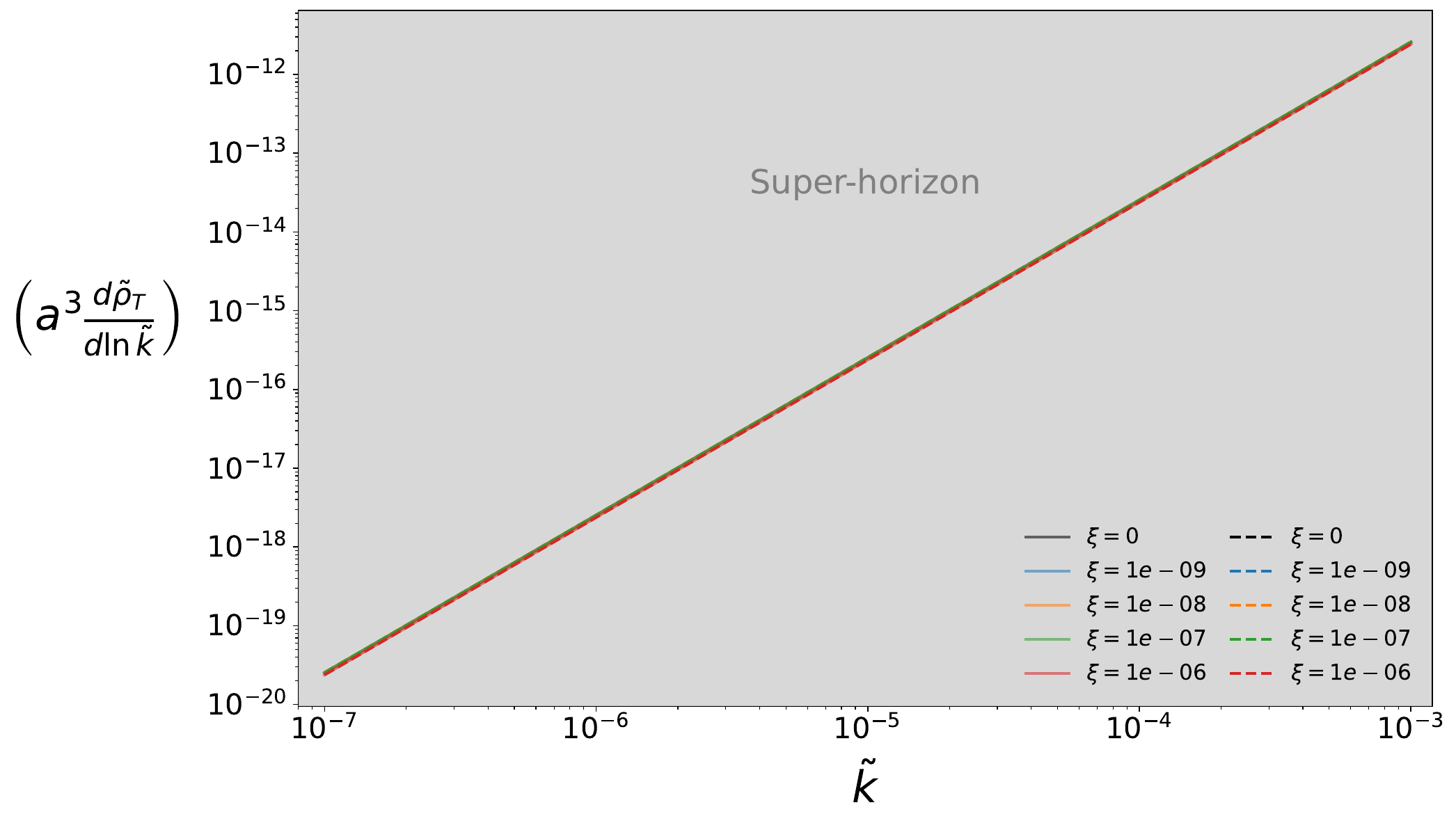}
\caption{
Spectral energy density of the transverse mode as a function of
the dimensionless wavenumber $\tilde{k}$, evaluated at $N=0$,
for the dS and qdS backgrounds and for
$0\leq\xi\leq10^{-6}$, with $\epsilon=10^{-3}$.
In the dS background, the effective mass squared is constant,
$\tilde{M}^{2}_{\mathrm{dS}}=\tilde{m}^{2}+12\xi$,
whereas in the qdS background it evolves as
$\tilde{M}^{2}_{\mathrm{qdS}}(N)
=\tilde{m}^{2}+6\xi(1+\alpha)e^{-2\epsilon N}$.
At the evaluation time $N=0$, the latter becomes
$\tilde{M}^{2}_{\mathrm{qdS}}(0)
=\tilde{m}^{2}+6\xi(1+\alpha)$.
For $\epsilon=10^{-3}$ and $\alpha=0.999$, the coefficient
$6(1+\alpha)=11.994$ is slightly smaller than the
corresponding coefficient $12$ in the dS background.
}
\label{fig:rho_T_vs_k_N0}
\end{figure}

\noindent
Figure~\ref{fig:rho_T_vs_k_N0} shows the transverse-mode spectral energy density, \(a^3 d\tilde{\rho}_T/d\ln\tilde{k}\), at \(N=0\) as a function of \(\tilde{k}\). Solid and dashed curves represent the dS and qdS backgrounds, respectively, for different values of \(\xi\). The range \(10^{-7}\leq\tilde{k}\leq10^{-3}\) corresponds to super-horizon modes.

\noindent

\begin{figure}[H]
\centering
\includegraphics[width=1.0\textwidth]{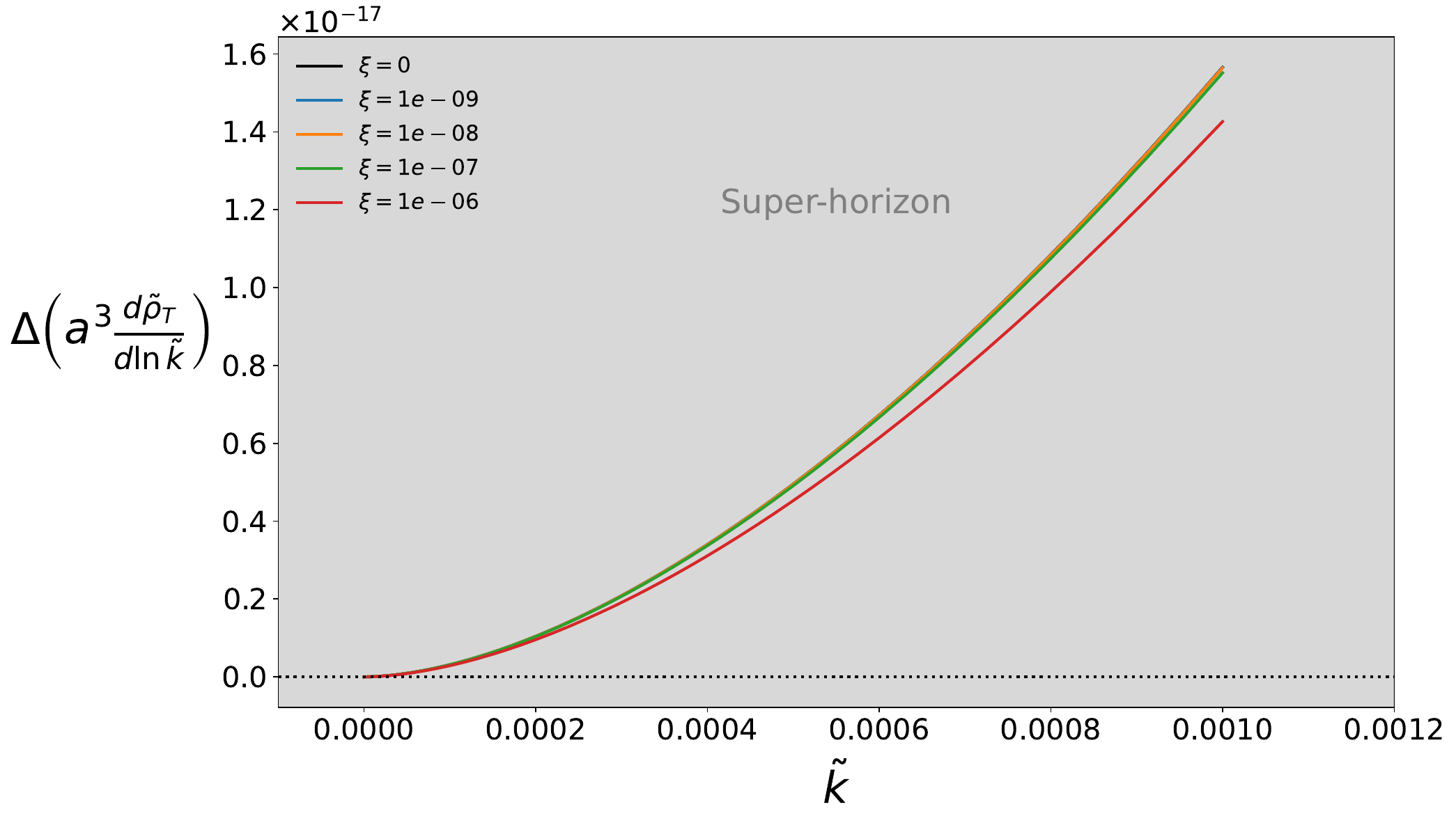}
\caption{
Difference between the transverse-mode spectral energy densities
in the qdS and dS backgrounds at $N=0$, as defined in
Eq.~\eqref{eq:difference}, for
$0\leq\xi\leq10^{-6}$ and $\epsilon=10^{-3}$.
}
\label{fig:difference_rho_T_vs_k_N0}
\end{figure}

\noindent
Figure~\ref{fig:difference_rho_T_vs_k_N0} shows the signed difference between the qdS and dS spectra, as defined in Eq.~\eqref{eq:difference}. Positive (negative) values indicate a larger (smaller) spectral energy density in the qdS background.

\subsubsection{Longitudinal Mode}

In this section, we investigate the dependence of the longitudinal-mode
spectral energy density, given by Eq.~\eqref{eq:spectral_rho_L_compact},
on the non-minimal coupling parameter $\xi$ in both dS and qdS
backgrounds. The analysis is performed over the range of dimensionless
wavenumbers $10^{-7}\leq\tilde{k}\leq10^{-3}$, with the spectral
energy densities evaluated at $N=0$, corresponding to the transition
from inflation to radiation domination.

\noindent
To quantify the impact of slow-roll corrections, we calculate the
difference between the qdS and dS longitudinal-mode spectral energy
densities, defined in Eq.~\eqref{eq:difference}.
This comparison allows us to assess how the departure from the dS
background to the qdS background affects the longitudinal-mode
spectrum across different wavenumber scales and values of $\xi$.

\begin{figure}[H]
\centering
\includegraphics[width=1.0\textwidth]{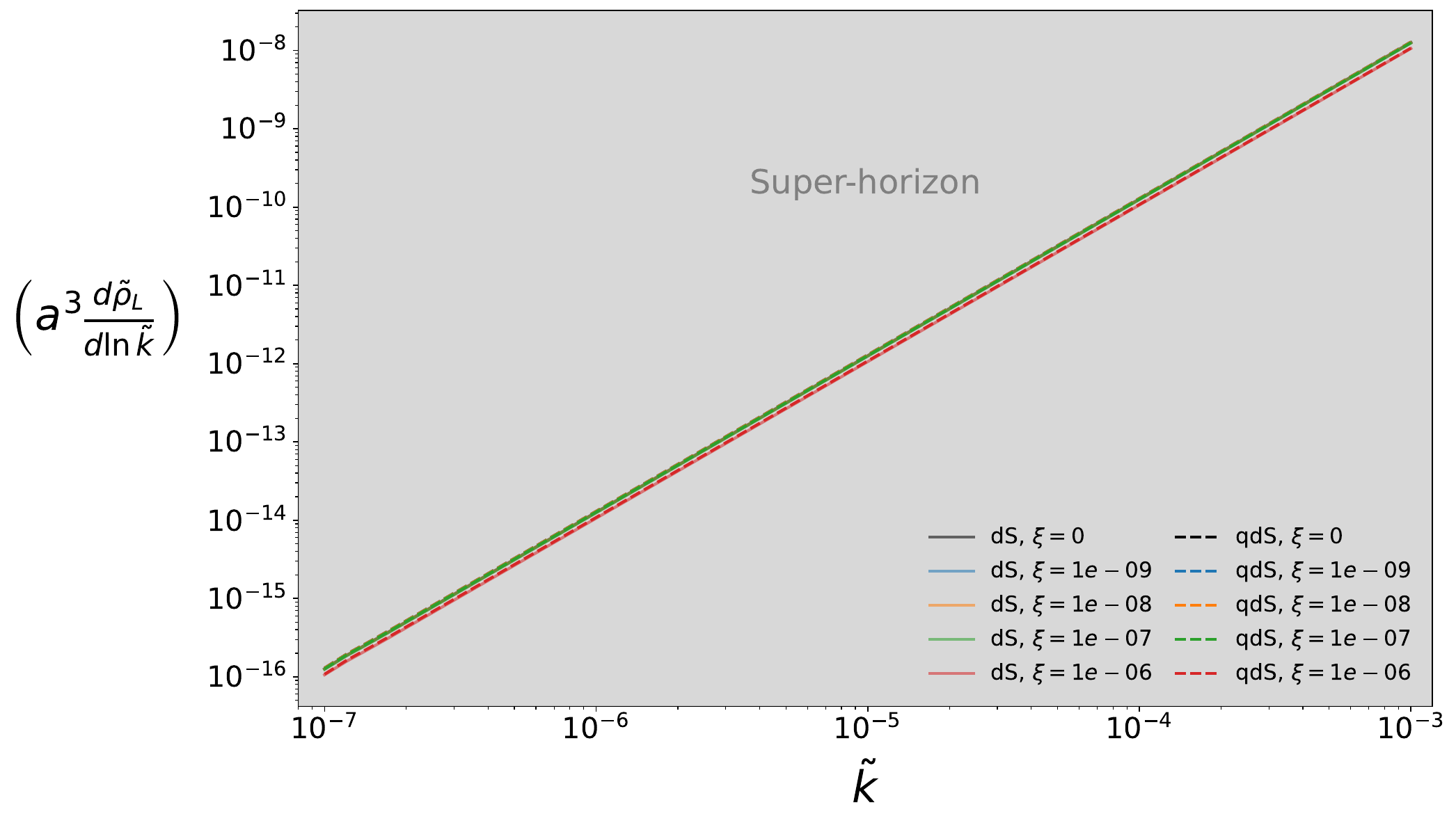}
\caption{
Spectral energy density of the longitudinal mode as a function of
the dimensionless wavenumber $\tilde{k}$, evaluated at $N=0$,
for the dS and qdS backgrounds, with
$0\leq\xi\leq10^{-6}$ and $\epsilon=10^{-3}$.
}
\label{fig:rho_L_vs_k_N0}
\end{figure}

\noindent
Figure~\ref{fig:rho_L_vs_k_N0} presents the longitudinal spectral
energy density, $a^3 d\tilde{\rho}_L/d\ln\tilde{k}$, evaluated
at $N=0$ as a function of the dimensionless comoving wavenumber
$\tilde{k}$ for different values of the non-minimal coupling
parameter $\xi$. The solid and dashed curves correspond to
the dS and qdS backgrounds,
respectively. The momentum range
$10^{-7}\leq\tilde{k}\leq10^{-3}$ probes the longitudinal
spectrum across different super-horizon scales. Comparing
the curves for different values of $\xi$ illustrates how the
non-minimal curvature coupling modifies the spectral energy
density in each background, while comparing the solid and
dashed curves highlights the effects of deviations from
exact de Sitter expansion.

\begin{figure}[H]
\centering
\includegraphics[width=1.0\textwidth]{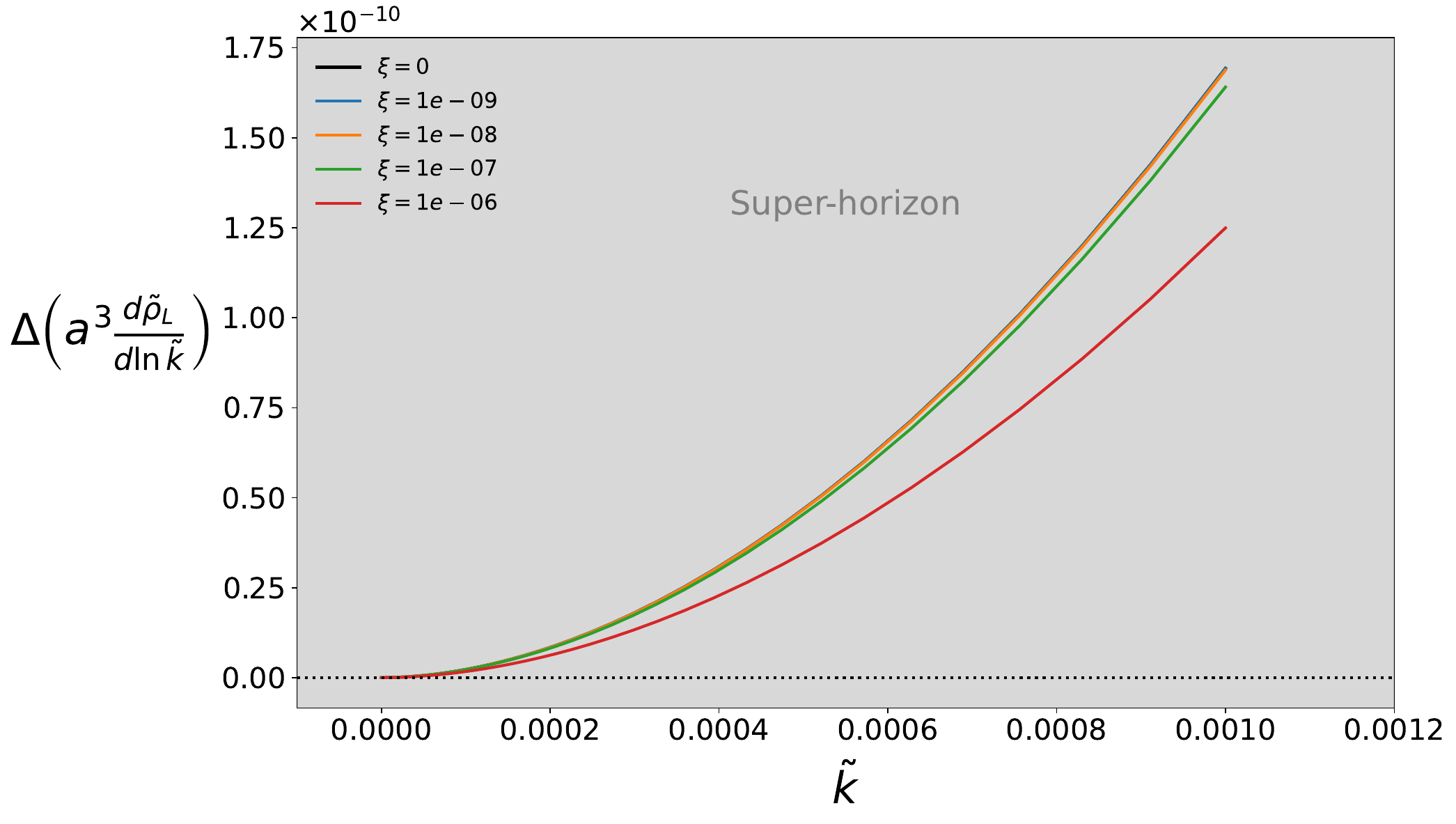}
\caption{
Difference between the longitudinal-mode spectral energy densities
in the qdS and dS backgrounds at $N=0$, as defined in
Eq.~\eqref{eq:difference}, for
$0\leq\xi\leq10^{-6}$ and $\epsilon=10^{-3}$.
}
\label{fig:difference_rho_L_vs_k_N0}
\end{figure}

\noindent
Figure~\ref{fig:difference_rho_L_vs_k_N0} shows the difference between the longitudinal spectral energy densities in the qdS and dS backgrounds, $\Delta_L$, defined in Eq.~\eqref{eq:difference}, as a function of the dimensionless comoving wavenumber $\tilde{k}$ for several values of the non-minimal coupling $\xi$. The dependence on $\tilde{k}$ shows how the two backgrounds differ across the considered momentum range, and the dependence on $\xi$ shows how the curvature coupling affects this difference. Together, they quantify the sensitivity of the longitudinal spectrum to deviations from exact de Sitter expansion.

\noindent
Further details of these effects are provided in Appendices~\ref{app:C} and~\ref{app:D}.

\subsection{Effective Frequency of the Longitudinal Mode}

\begin{figure}[H]
\centering
\includegraphics[width=1.0\textwidth]{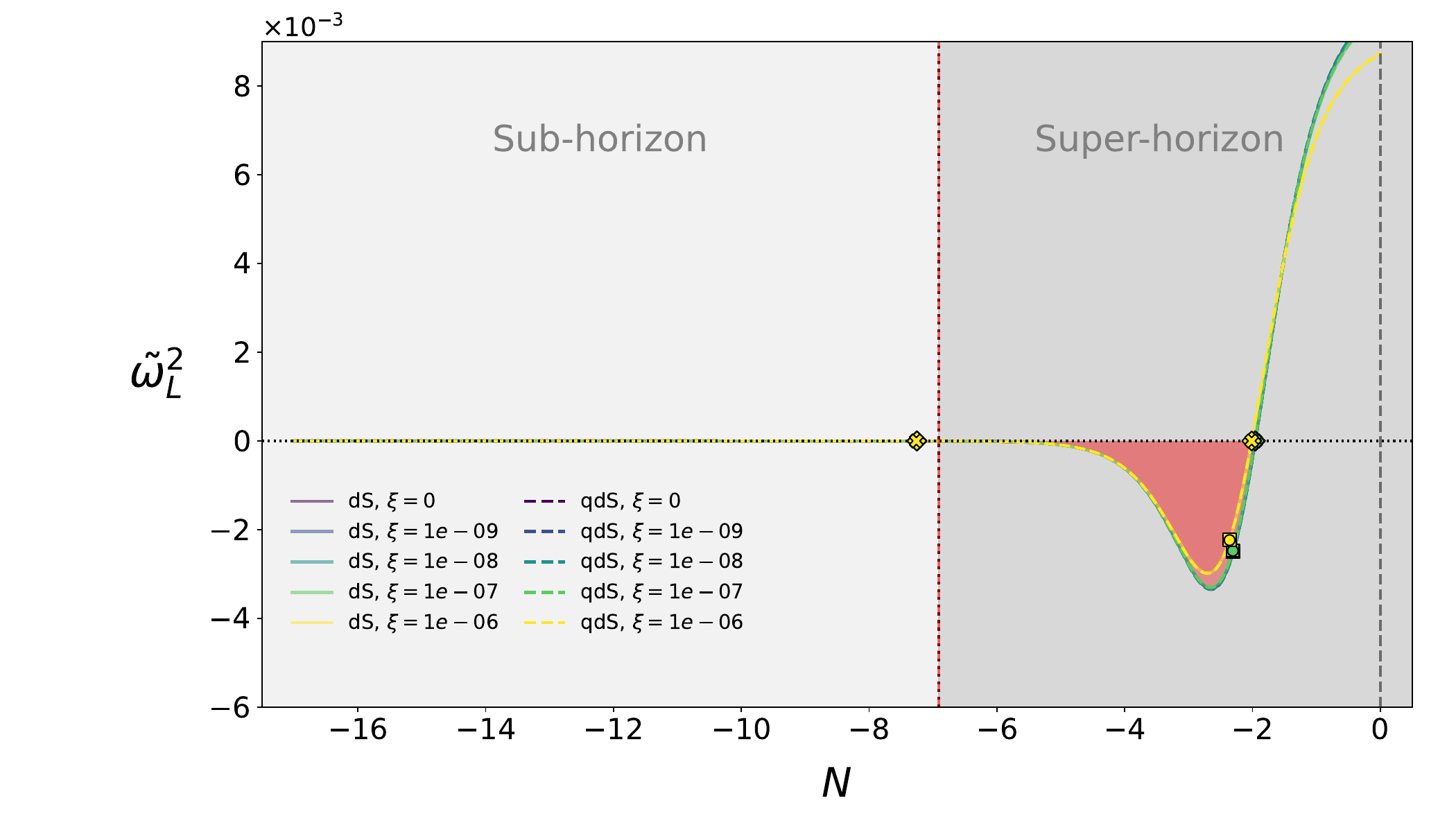}
\caption{Squared effective frequency of the longitudinal mode,
$\tilde{\omega}_L^2$, in the dS and qdS backgrounds for the range
$0 \leq \xi \leq 10^{-6}$ with $\epsilon = 10^{-3}$.
The effective mass is given by
$\tilde{M}^{2} = \tilde{m}^{2} + 12\xi$ in dS and
$\tilde{M}^{2} = \tilde{m}^{2} + 6\xi(1+\alpha)e^{-2\epsilon N}$ in qdS.
The vertical lines indicate the characteristic times during the
mode evolution: the horizon-crossing time in the dS background,
$N = \ln(\tilde{k}) = -6.90775$, the horizon-crossing time in the
qdS background, $N = \ln(\tilde{k})/\alpha = -6.91466$, and the
transition from the inflationary phase to the RD epoch at $N=0$.}
\label{fig:omega_L2_Total}
\end{figure}

\begin{figure}[H]
\centering
\includegraphics[width=1.0\textwidth]{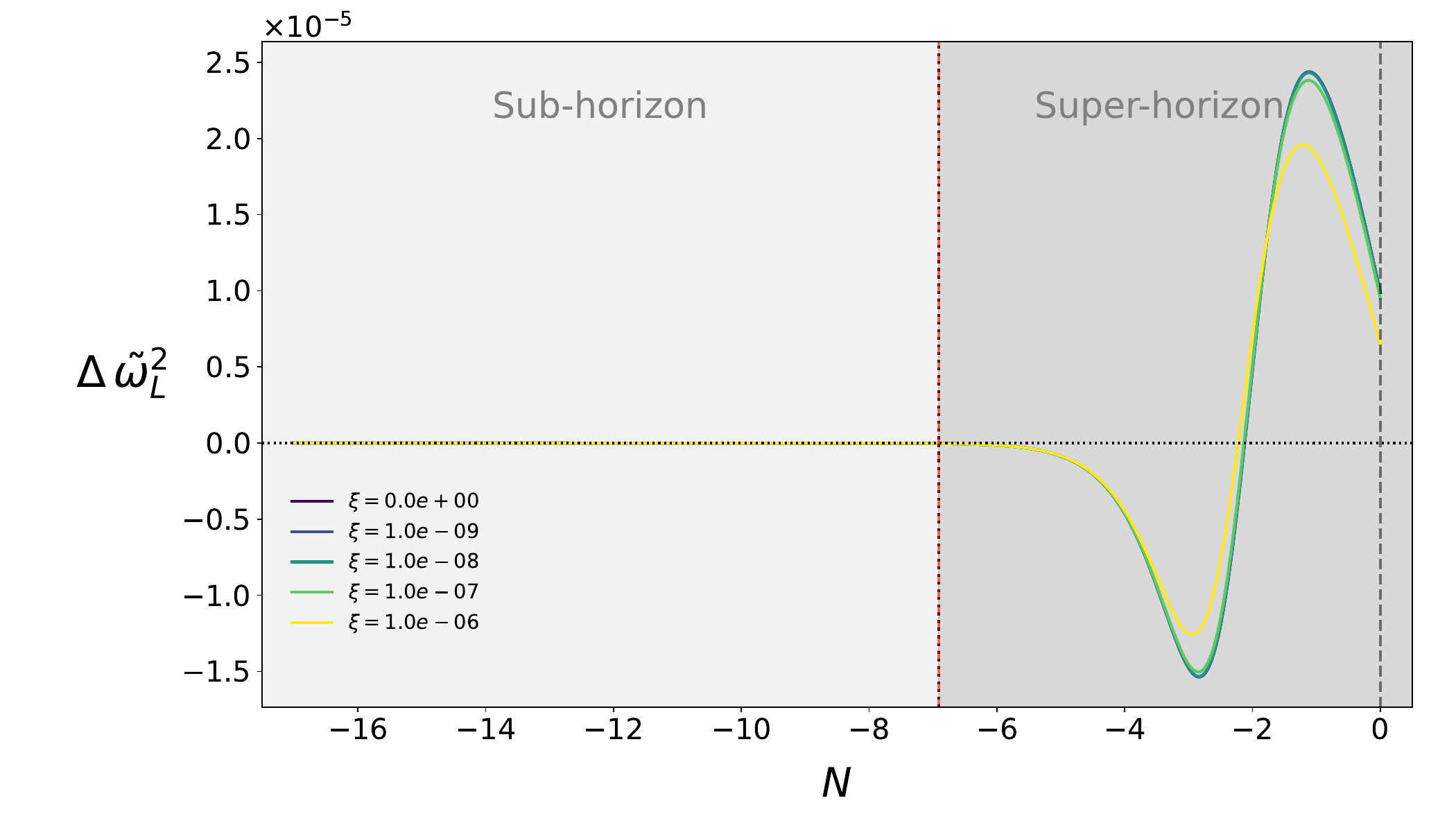}
\caption{Difference between the squared longitudinal-mode effective
frequencies in the qdS and dS backgrounds, defined as
$\Delta\tilde{\omega}_L^2=
\tilde{\omega}_{L,\mathrm{qdS}}^2-
\tilde{\omega}_{L,\mathrm{dS}}^2$
in Eq.~\eqref{eq:difference}, for
$0\leq\xi\leq10^{-6}$ and $\epsilon=10^{-3}$.
The dS frequency is linearly interpolated onto the qdS time grid.
The effective mass is given by
$\tilde{M}^{2}=\tilde{m}^{2}+12\xi$ for dS and
$\tilde{M}^{2}=\tilde{m}^{2}+6\xi(1+\alpha)e^{-2\epsilon N}$
for qdS.
The vertical lines indicate the horizon-crossing times in the dS
and qdS backgrounds, respectively,
$\ln(\tilde{k})=-6.90775$ and
$\ln(\tilde{k})/\alpha=-6.91466$, and the transition
from inflation to the RD epoch at $N=0$.}
\label{fig:difference_omega_L2_Total}
\end{figure}

In this subsection, we investigate the effective frequency of the longitudinal mode and its evolution in both dS and qdS backgrounds for different values of the non-minimal coupling parameter $\xi$. We plot the normalized frequency defined in Eq.~\eqref{eq:omega_L2_tilde} as a function of $N$ in Figs.~\ref{fig:omega_L2_Total} and \ref{fig:difference_omega_L2_Total}. We also present the corresponding crossing points and roots in Tables~\ref{tab:omegaL_crossing} and \ref{tab:omegaL2_roots}, respectively. The longitudinal sector exhibits non-trivial behavior due to the effective mass and additional contributions arising from the curvature coupling. We analyze the dependence of the effective frequency on $\xi$ and compare its evolution in the two inflationary backgrounds.

\noindent
The squared effective frequency of the longitudinal mode in the
minimal coupling case ($\xi=0$) is given by
Eq.~\eqref{eq:omega_L2}, whereas its expression in the non-minimal
coupling case ($\xi\neq0$) is given by
Eq.~\eqref{eq:omega_L2_eff}.

\noindent
Figure~\ref{fig:omega_L2_Total} shows the evolution of the squared
effective frequency of the longitudinal mode for different values
of the non-minimal coupling parameter $\xi$ in both dS and qdS
backgrounds. For minimal coupling ($\xi=0$), the effective mass
squared is constant and identical in both backgrounds,
$\tilde{M}^2=\tilde{m}^2$. Nevertheless, the corresponding
frequency evolutions are not identical because of the difference
between the dS and qdS background evolution. The resulting
deviation is small, as also reflected in
Table~\ref{tab:omegaL_crossing}.

\noindent
As $\xi$ increases, the curvature contribution modifies the
effective mass. In particular, the effective mass remains constant
in the dS background, whereas it becomes time-dependent in the
qdS background. Consequently, the difference between the dS and
qdS frequency evolutions becomes more pronounced for the larger
values of $\xi$ considered here, illustrating the influence of
the non-minimal curvature coupling on the longitudinal-mode
dynamics.

\label{subsec:longitudinal_frequency_stability}

\noindent
We investigate the evolution of the squared longitudinal frequency, 
$\omega_L^2$, in dS and qdS backgrounds, using the normalized quantity 
defined in Eq.~\eqref{eq:omega_L2_tilde}. 
In our backgrounds,
\begin{equation}
    H(N)=H_{\rm end}e^{-\epsilon N},
    \label{eq:background_frequency_scales}
\end{equation}
where $\alpha$ is defined in Eq.~\eqref{eq:alpha_definition}.
Here and throughout the paper, $H$ denotes the time-dependent
Hubble parameter $H(N)$, unless otherwise specified.
We use two different normalizations of the longitudinal frequency. The first
is defined in Eq.~\eqref{eq:omega_L2_tilde} and uses $H_{\rm end}$. The
second normalizes with the time-dependent Hubble rate $H(N)$,
\begin{equation}
\tilde{\omega}_{L,H}^2
\equiv\frac{\omega_L^2}{H(N)^2}
=\tilde{\omega}_L^2 e^{2\epsilon N}.
\label{eq:frequency_normalization_relations}
\end{equation}
The two normalizations differ by the factor $e^{2\epsilon N}$ and coincide in
the dS limit $\epsilon\to0$. The normalization used is specified explicitly in
each figure and table.

\noindent
Table~\ref{tab:omegaL_crossing} summarizes the e-fold number at which the mass-crossing condition, Eq.~(\ref{eq:mass_crossing}), is satisfied, together with the corresponding values of the effective frequency squared of the longitudinal mode, $\tilde{\omega}_{L}^{2}$, for different values of the non-minimal coupling parameter $\xi$ in both the dS and qdS backgrounds.

\begin{table}[H]
\centering
\small
\setlength{\tabcolsep}{4pt}

\caption{The squared longitudinal frequency,
$\tilde{\omega}_{L}^{2}=\omega_L^2/H_{\rm end}^{2}$,
evaluated at the effective-mass crossing condition $M=k/a$
for different values of the non-minimal coupling parameter
$\xi$ in the dS and qdS backgrounds.}
\label{tab:omegaL_crossing}

\begin{tabular}{|c|c|c|c|c|}
\hline
$\xi$ &
$N_{M=k/a}^{\rm dS}$ &
$\left(\tilde{\omega}_{L}^{2}\right)_{\mathrm{dS}}$ &
$N_{M=k/a}^{\rm qdS}$ &
$\left(\tilde{\omega}_{L}^{2}\right)_{\mathrm{qdS}}$
\\
\hline

$0$
& $-2.30259$
& $-2.49797\times10^{-3}$
& $-2.30259$
& $-2.50449\times10^{-3}$
\\

$1\times10^{-9}$
& $-2.30265$
& $-2.49767\times10^{-3}$
& $-2.30265$
& $-2.50418\times10^{-3}$
\\

$1\times10^{-8}$
& $-2.30318$
& $-2.49498\times10^{-3}$
& $-2.30319$
& $-2.50148\times10^{-3}$
\\

$1\times10^{-7}$
& $-2.30855$
& $-2.46835\times10^{-3}$
& $-2.30857$
& $-2.47470\times10^{-3}$
\\

$1\times10^{-6}$
& $-2.35925$
& $-2.23014\times10^{-3}$
& $-2.35948$
& $-2.23520\times10^{-3}$
\\

\hline
\end{tabular}
\end{table}
\noindent
Table~\ref{tab:omegaL2_roots} lists the zero-crossing points of
$\tilde{\omega}_L^2(N,\xi)$, which delimit the intervals in $N$
where the squared longitudinal frequency is negative, for both
dS and qdS backgrounds and for
$0\leq\xi\leq10^{-6}$.

\begin{table}[H]
\centering
\small
\setlength{\tabcolsep}{4pt}
\renewcommand{\arraystretch}{1.05}

\caption{The two roots, $N_{\mathrm{dS},1}$ and $N_{\mathrm{dS},2}$,
of $\tilde{\omega}_L^2(N,\xi)=0$ in the dS background, and
$N_{\mathrm{qdS},1}$ and $N_{\mathrm{qdS},2}$ in the qdS background,
for different values of the non-minimal coupling parameter $\xi$.
The first and second roots denote the earlier and later zero crossings
in $N$, respectively.}
\label{tab:omegaL2_roots}

\begin{tabular}{|c|c|c|c|c|}
\hline
$\xi$ &
$N_{\mathrm{dS},1}$ &
$N_{\mathrm{dS},2}$ &
$N_{\mathrm{qdS},1}$ &
$N_{\mathrm{qdS},2}$
\\
\hline

$0$
& $-7.254246$
& $-1.956350$
& $-7.261255$
& $-1.957098$
\\

$1\times10^{-9}$
& $-7.254246$
& $-1.956410$
& $-7.261255$
& $-1.957159$
\\

$1\times10^{-8}$
& $-7.254246$
& $-1.956950$
& $-7.261254$
& $-1.957700$
\\

$1\times10^{-7}$
& $-7.254245$
& $-1.962319$
& $-7.261245$
& $-1.963078$
\\

$1\times10^{-6}$
& $-7.254235$
& $-2.013055$
& $-7.261163$
& $-2.013912$
\\

\hline
\end{tabular}
\end{table}
\noindent
Figure~\ref{fig:difference_omega_L2_Total} shows the difference
between the squared longitudinal frequencies in the qdS and dS
backgrounds for different values of the non-minimal coupling
parameter $\xi$. The difference is defined in
Eq.~\eqref{eq:difference}, where the dS frequency is linearly
interpolated onto the qdS time grid before evaluating the difference.

\section{Conclusion}

In this work, we investigated the evolution of a spectator massive vector
field in a spatially flat FLRW universe, considering both minimal and
non-minimal couplings to gravity. We analyzed the transverse and
longitudinal polarization modes separately and tracked their evolution
in exact dS, qdS, and RD backgrounds. Our main objective was to quantify the impact of the
Ricci-scalar coupling and slow-roll corrections on the spectral energy
densities and the dynamical evolution of vector-field fluctuations
during inflation and across the transition to the RD era.

\noindent
The inflationary and RD phases are connected through an idealized instantaneous transition at N=0. We
impose continuity of the scale factor and Hubble parameter, together with the appropriate matching conditions for the
transverse and longitudinal field variables and their canonical momenta.

\noindent
For the transverse sector, our numerical results show that when evaluated at a fixed common cosmological time, the spectral
energy density generally decreases with increasing non-minimal coupling
parameter $\xi$ in both dS and qdS backgrounds. This behavior results
from the curvature-induced contribution to the effective mass, which
increases the mass scale of the vector field and suppresses the
production of transverse fluctuations. In the minimally coupled limit,
the dS and qdS solutions remain very close, whereas increasing $\xi$
enhances the difference between the two backgrounds due to the explicit
time dependence of the effective mass in the qdS case. The comparison
of the absolute and normalized differences confirms that slow-roll
corrections introduce relatively small quantitative modifications,
although their effects become more visible near the end of inflation
and around the transition to radiation domination.

\noindent
The mass-crossing analysis shows that increasing $\xi$ shifts the
crossing epoch toward earlier e-folds in both backgrounds. At the
corresponding $\xi$-dependent mass-crossing time, the transverse-mode
spectral energy density increases monotonically with $\xi$, as shown
in Table~1. This behavior should be distinguished from the suppression
observed in spectra evaluated at a fixed common time: because the
mass-crossing condition is reached at progressively earlier epochs
as $\xi$ increases, the values in Table~1 correspond to different
times for different couplings. For nonzero $\xi$, the qdS values are
consistently slightly larger than the corresponding dS values,
indicating a small enhancement due to slow-roll corrections at the
mass-crossing epoch.

\noindent
Unlike the transverse modes, the longitudinal polarization is affected
by its nontrivial kinetic normalization, which makes its dynamics
particularly sensitive to the background evolution and the time
dependence of the effective frequency. Consequently, the longitudinal
mode provides additional information about the dynamical behavior of
the system through its effective frequency squared,
$\tilde{\omega}_L^2$. The comparison between dS and qdS backgrounds
shows that slow-roll corrections modify both the effective frequency
and the resulting longitudinal spectral energy density. These
differences arise from the combined effects of the time-dependent
effective mass and the nontrivial kinetic structure of the longitudinal
polarization. In particular, the longitudinal sector exhibits a more
complicated dependence on the background evolution than the transverse
sector, emphasizing the importance of treating the two polarizations
separately.

\noindent
For the numerical evolution, we imposed second-order adiabatic WKB
initial conditions corresponding to the Bunch--Davies positive-frequency
vacuum in the subhorizon regime.
The second-order adiabatic correction to the WKB frequency was
included using the first and second derivatives of the zeroth-order
frequency with respect to the e-folding variable $N$.

\noindent
The initial conditions were imposed at $N_{\rm ini}=-17$
in both the dS and qdS backgrounds. For $\tilde{k}=10^{-3}$,
this corresponds to
\begin{equation}
    \left.\frac{k}{aH}\right|_{\rm ini}^{\rm dS}
    =\tilde{k}e^{17}\simeq2.42\times10^4,
    \qquad
    \left.\frac{k}{aH}\right|_{\rm ini}^{\rm qdS}
    =\tilde{k}e^{17\alpha}\simeq2.38\times10^4.
    \label{eq:conclusion_wkb_initial_depth}
\end{equation}
The initial conditions were therefore specified deep inside the horizon in
both backgrounds, where the Bunch--Davies vacuum is well
defined~\cite{Baumann2009,Riotto2003}. The range of the non-minimal coupling
parameter was chosen in accordance with the theoretical constraint discussed
in Ref.~\cite{LebedevYoon2026},
\begin{equation}
    |\xi| \ll \frac{m^{2}}{6H^{2}}.
    \label{eq:conclusion_xi_constraint}
\end{equation}
This condition must be checked throughout the inflationary
interval, rather than only at the end of inflation.
Using $H(N)=H_{\rm end}e^{-\epsilon N}$ and
$\tilde{m}=m/H_{\rm end}$, the corresponding reference
scale becomes
\begin{equation}
    \frac{m^2}{6H^2(N)}
    =\frac{\tilde{m}^2}{6}e^{2\epsilon N}.
    \label{eq:conclusion_xi_reference_scale}
\end{equation}
For the dS background, $\epsilon=0$, and this scale remains
constant at $\tilde{m}^2/6\simeq1.67\times10^{-5}$.
For the qdS background, $\epsilon>0$, so the most restrictive
bound occurs at the initial time. With
$\tilde{m}=10^{-2}$, $\epsilon=10^{-3}$, and
$N_{\rm ini}=-17$, the minimum reference scale is
\begin{equation}
    \min_{N_{\rm ini}\leq N\leq0}
    \frac{m^2}{6H^2(N)}
    =
    \frac{\tilde{m}^2}{6}e^{2\epsilon N_{\rm ini}}
    \simeq1.61\times10^{-5}.
    \label{eq:conclusion_xi_reference_min}
\end{equation}
For the coupling range specified in Eq.~\eqref{eq:xi_range},
the largest value considered, $\xi=10^{-6}$, yields
\begin{equation}
    \max_{N_{\rm ini}\leq N\leq0}
    \frac{6|\xi|H^2(N)}{m^2}
    \simeq0.0621.
    \label{eq:conclusion_xi_hierarchy}
\end{equation}
The largest coupling therefore amounts to approximately
$6.2\%$ of the most restrictive reference scale, with
smaller couplings providing a correspondingly stronger
hierarchy.

\noindent
We also examined the evolution across the instantaneous transition
from inflation to RD. The transverse and longitudinal solutions were
matched by imposing continuity of the corresponding field variables
and canonical momenta. In particular, the longitudinal matching
conditions require continuity of $A_L$ and $K_L A_L'$, rather than
of $A_L'$ itself, as detailed in Appendix~C.4. These conditions
account for the discontinuity in the longitudinal kinetic
normalization associated with the change in the curvature-dependent
effective mass at the transition. These conditions account for the nontrivial kinetic normalization of the
longitudinal polarization and ensure a consistent connection between
the inflationary and RD solutions.

\noindent
During RD, the disappearance of the curvature-induced contribution
means that the subsequent evolution is controlled by the bare vector
mass and the expansion of the universe. The post-inflationary
evolution therefore provides a useful way to distinguish effects
originating from the non-minimal coupling during inflation from those
associated with the subsequent massive-field dynamics. In particular,
the spectral energy densities retain information about the
inflationary evolution through the mode functions and their matching
data at the transition.

\noindent
Overall, our results demonstrate that the transverse and longitudinal
polarizations respond differently to non-minimal curvature coupling and
slow-roll corrections. While the transverse sector primarily exhibits a
coupling-dependent suppression at fixed evaluation time, the longitudinal
sector is additionally sensitive to its kinetic normalization and
effective-frequency structure. The comparison between dS and qdS backgrounds
shows that slow-roll corrections generally produce quantitative modifications
to the mode evolution and spectral energy densities, while the transition to
RD removes the direct curvature-induced contribution to the vector mass.
Together, these results provide a unified numerical investigation of the
inflationary and post-inflationary evolution of a massive spectator vector
field with non-minimal curvature coupling. They also complement existing
studies of generalized Proca
theories~\cite{DeFelice2016Cosmology,DeFelice2016Effective,Heisenberg2016,Heisenberg2017,Heisenberg2018}
and of the stability of the longitudinal
mode~\cite{Jimenez2012}.

\appendix
\numberwithin{equation}{section}

\section{Computation of the non-minimal part of the energy density}
\label{app:A}

The Proca contribution to the energy-momentum tensor is given by

\begin{equation}
T_{\mu\nu}^{(\mathrm{Proca})}
=
F_{\mu\lambda}F_{\nu\beta}g^{\lambda\beta}
+
m^2 A_\mu A_\nu
-
g_{\mu\nu}
\left[
\frac{1}{4}F^2
+
\frac{1}{2}m^2 A^2
\right],
\end{equation}
\noindent
while the contribution arising from the non-minimal curvature coupling is

\begin{equation}
T_{\mu\nu}^{(\xi)}
=
\xi
\left[
G_{\mu\nu}X
+
R A_{\mu}A_{\nu}
+
g_{\mu\nu}\Box X
-
\nabla_{\mu}\nabla_{\nu}X
\right],
\label{eq:T_mu_nu_xi}
\end{equation}
where we have defined
\begin{equation}
X \equiv A_{\alpha}A^{\alpha}.
\end{equation}
In the spatially flat FLRW background,

\begin{equation}
X
=
A_{\mu}A^{\mu}
=
\frac{-A_{0}^{2}+|\mathbf{A}|^{2}}{a^{2}},
\end{equation}
With the metric~\eqref{eq:FLRW_metric}, the relevant components for this background are
\begin{equation}
G_{00}
=
3\mathcal H^{2},
\qquad
\left(
g_{00}\Box
-
\nabla_{0}\nabla_{0}
\right)X
=
3\mathcal HX'
-
\nabla^{2}X .
\label{eq:background_components}
\end{equation}
Therefore, the temporal component of the energy-momentum tensor becomes

\begin{equation}
T_{00}
=
T_{00}^{\rm(Proca)}
+
\xi
\left(
3\mathcal H^{2}X
+
RA_{0}^{2}
+
3\mathcal HX'
-
\nabla^{2}X
\right),
\label{eq:T_00_nonminimal_total}
\end{equation}
where the Proca contribution is given by

\begin{equation}
T_{00}^{(\mathrm{Proca})}
=
\frac{1}{2a^2}
\left[
F_{0i}F_{0i}
+
\frac{1}{2}F_{ij}F_{ij}
+
m^2 a^2
\left(
A_0^2+A_iA_i
\right)
\right].
\label{eq:T_00_Proca}
\end{equation}
The derivative terms appearing in the non-minimal contribution are

\begin{equation}
X'
=
\frac{1}{a^{2}}
\left[
-2A_{0}A_{0}'
+
2\mathbf A\cdot\mathbf A'
-
2\mathcal H
\left(
-A_{0}^{2}
+
|\mathbf A|^{2}
\right)
\right],
\label{eq:X_prime_Proca}
\end{equation}
and

\begin{equation}
\nabla^{2}X
=
\frac{2}{a^{2}}
\Big[
-A_{0}\nabla^{2}A_{0}
-
|\nabla A_{0}|^{2}
+\mathbf A\cdot\nabla^{2}\mathbf A
+
|\nabla A_{i}|^{2}
\Big].
\label{eq:Grad_X_Proca}
\end{equation}
Finally, the energy density is obtained from

\begin{equation}
\rho
=
-g^{00}T_{00}
=
\frac{T_{00}}{a^{2}},
\label{eq:rho_compact}
\end{equation}
The energy density can be obtained
\begin{equation}
\begin{aligned}
\rho
=&\frac{1}{a^{2}}
\Bigg\{
\frac{1}{2a^{2}}
\left[
F_{0i}F_{0i}
+\frac{1}{2}F_{ij}F_{ij}
+m^{2}a^{2}
\left(A_{0}^{2}+A_iA_i\right)
\right]
\\
&+\xi
\Bigg[
3\mathcal{H}^{2}X
+R A_{0}^{2}
+\frac{3\mathcal{H}}{a^{2}}
\left(
-2A_{0}A_{0}^{\prime}
+2\mathbf{A}\cdot\mathbf{A}^{\prime}
-2\mathcal{H}
\left(-A_{0}^{2}+|\mathbf{A}|^{2}\right)
\right)
\\
&\qquad
-\frac{2}{a^{2}}
\left(
-A_{0}\nabla^{2}A_{0}
-|\nabla A_{0}|^{2}
+\mathbf{A}\cdot\nabla^{2}\mathbf{A} + |\nabla A_i|^{2}
\right)
\Bigg]
\Bigg\}.
\end{aligned}
\end{equation}
which, in Fourier space, reads

\begin{equation}
\rho
=
\rho_{\rm compact}
+
\frac{\xi}{2a^{4}}
\int
\frac{d^{3}k\,d^{3}p}{(2\pi)^3}
e^{i\mathbf{x}\cdot(\mathbf{k}+\mathbf{p})}
\left[
\Delta_{\mathcal H}(\mathbf{k},\mathbf{p})
+
\Delta_{\nabla}(\mathbf{k},\mathbf{p}) + \Delta
\right]
\end{equation}
where
\begin{equation}
\begin{aligned}
\rho_{\rm compact}
=&
\frac{1}{2a^{4}}
\int
\frac{d^{3}k\,d^{3}p}{(2\pi)^3}
e^{i\mathbf{x}\cdot(\mathbf{k}+\mathbf{p})}
\Bigg\{
\\
&
\left[
\mathbf A_T'(\mathbf k)
+\hat{\mathbf k}A_L'(\mathbf k)
-i\mathbf k A_0(\mathbf k)
\right]
\cdot
\left[
\mathbf A_T'(\mathbf p)
+\hat{\mathbf p}A_L'(\mathbf p)
-i\mathbf p A_0(\mathbf p)
\right]
\\
&
+
\left[
i\mathbf k\times\mathbf A_T(\mathbf k)
\right]
\cdot
\left[
i\mathbf p\times\mathbf A_T(\mathbf p)
\right]
\\
&
+
M^{2}a^{2}
\Big[
A_0(\mathbf k)A_0(\mathbf p)
+
\mathbf A(\mathbf k)\cdot\mathbf A(\mathbf p)
\Big]
\Bigg\}.
\end{aligned}
\end{equation}
with
\begin{equation}
\mathbf A(\mathbf k)
=
\mathbf A_T(\mathbf k)
+\hat{\mathbf k}A_L(\mathbf k),
\end{equation}
The additional non-minimal derivative contribution is
\begin{equation}
\begin{aligned}
\Delta_{\mathcal H}(\mathbf{k},\mathbf{p})
=&
6\mathcal H
\Bigg\{
-A_0'(\mathbf{k})A_0(\mathbf{p})
-A_0(\mathbf{k})A_0'(\mathbf{p})
\\
&
+
\mathbf A'(\mathbf{k})\cdot\mathbf A(\mathbf p)
+
\mathbf A(\mathbf k)\cdot\mathbf A'(\mathbf p)
\\
&
-
2\mathcal H
\left[
-A_0(\mathbf{k})A_0(\mathbf{p})
+
\mathbf A(\mathbf{k})\cdot\mathbf A(\mathbf p)
\right]
\Bigg\},
\end{aligned}
\end{equation}
\begin{equation}
\Delta_{\nabla}(\mathbf{k},\mathbf{p})
=
-2\,|\mathbf{k} + \mathbf{p}|^2
\left[
A_0(\mathbf{k})A_0(\mathbf{p})
-
\mathbf{A}(\mathbf{k})\cdot\mathbf{A}(\mathbf{p})
\right],
\label{eq:delta_delta}
\end{equation}
and
\begin{equation}
\Delta = \left(a^2 R - 6 \mathcal{H}^2\right)\left[
A_0(\mathbf{k})A_0(\mathbf p)
-
\mathbf A(\mathbf{k})\cdot\mathbf A(\mathbf p)
\right].
\end{equation}
Finally, the constraint is the following.
\begin{equation}
A_0(\mathbf{k})
=
-\frac{ik}
{k^{2}+M^{2}a^{2}}
A_L'(\mathbf{k}).
\end{equation}

\noindent
The fields appearing in the previous expressions are quantum operators. 
The energy density is obtained by taking the vacuum expectation value.
Using the Fourier expansion of the transverse and longitudinal modes,

\begin{equation}
\langle A'_{\lambda}(\mathbf{k}) A_{\lambda}(\mathbf{p}) \rangle
=
\delta^{(3)}(\mathbf{k}+\mathbf{p})\,
A'_{\lambda}(\mathbf{k})A_{\lambda}^*(\mathbf{k}) 
\label{eq:expect}.
\end{equation}
After integration, including the Hermitian-conjugate term,
\begin{equation}
\left\langle
A'_{\lambda} A_{\lambda}
+
A_{\lambda} A'_{\lambda}
\right\rangle
=
2\,\mathrm{Re}\!\left[
A'_{\lambda} A_{\lambda}^{*}
\right].
\end{equation}
Also,
\begin{equation}
\left\langle
A'_0 A_0 + A_0 A'_0
\right\rangle
=
2 \, \mathrm{Re}\!\left[
A'_0 A_0^*
\right].
\end{equation}
From Eq.~\eqref{eq:expect}, we have
\begin{equation}
\mathbf{p} = -\mathbf{k}.
\end{equation}
Thus, in Eq.~\eqref{eq:delta_delta},
\begin{equation}
\left|\mathbf{k}+\mathbf{p}\right|^2 = 0.
\end{equation}
Therefore,
\begin{equation}
\left\langle \Delta_{\nabla} \right\rangle = 0.
\end{equation}
However, $\Delta_{\mathcal H}$ and $\Delta$ do not vanish.
	\begin{equation}
		\left\langle \rho \right\rangle
		=
		\left\langle \rho_{\mathrm{compact}} \right\rangle
		+
        \left\langle \rho_{\mathrm{non\text{-}minimal}} \right\rangle.
	\end{equation}
\begin{equation}
	\left\langle \rho_{\mathrm{non\text{-}minimal}} \right\rangle
	=
	\frac{\xi}{2a^4}
	\int \frac{d^3k}{(2\pi)^3}
	\left[
	\left\langle \Delta_{\mathcal H} \right\rangle
	+
	\left\langle \Delta \right\rangle
	\right].
\end{equation}
	After expressing the mode functions in terms of correlators and setting
	$\mathbf{p}=-\mathbf{k}$, we obtain
	\begin{equation}
		\left\langle \Delta_{\mathcal H} \right\rangle
		=
		12\mathcal H
		\left[
		\sum_i
		\mathrm{Re}\!\left(A'_i A_i^*\right)
		-
		\mathrm{Re}\!\left(A'_0 A_0^*\right)
		+
		\mathcal H
		\left(
		|A_0|^2-\sum_i |A_i|^2
		\right)
		\right].
	\end{equation}
	\begin{equation}
		\left\langle \Delta \right\rangle
		=
		\left(a^2R-6\mathcal{H}^2\right)
		\left(
		|A_0|^2-\sum_i |A_i|^2
		\right).
	\end{equation}
	\begin{equation}
		\begin{aligned}
			\left\langle
			\Delta_{\mathcal H}+\Delta
			\right\rangle
			={}&
			12\mathcal H
			\left[
			\sum_i
			\mathrm{Re}\!\left(A'_i A_i^*\right)
			-
			\mathrm{Re}\!\left(A'_0 A_0^*\right)
			\right]
			\\
			&+
			\left(a^2R+6\mathcal H^2\right)
			\left(
			|A_0|^2-\sum_i |A_i|^2
			\right).
		\end{aligned}
	\end{equation}

\clearpage
\section{Numerical Solutions for the Transverse Mode in dS, qdS, and RD Epochs}
\label{app:B}

In this appendix we describe the numerical solution of the transverse mode
during dS and qdS inflation, and its subsequent
evolution through the instantaneous transition to the RD
epoch. The mode function obeys
\begin{equation}
  A_{T}''(\tau)+\left(k^{2}+M^{2}a^{2}(\tau)\right)A_{T}(\tau)=0 ,
  \label{eq:transverse_eom}
\end{equation}
where primes denote derivatives with respect to conformal time. We define the
rescaled variable
\begin{equation}
  \tilde{A}_T(\tau)=\sqrt{2k}\,e^{ik\tau}A_T(\tau),
  \label{eq:tilde_A_T_transform}
\end{equation}
which removes the rapidly oscillating phase deep inside the horizon.
Throughout, tildes denote quantities in units of $H_{\rm end}$, i.e.
$\tilde k=k/H_{\rm end}$ and $\tilde M=M/H_{\rm end}$, and the scale factor is
normalized to $a_{\rm end}=1$ at the end of inflation. The conformal time in each
epoch is given by Eqs.~\eqref{eq:tau_N} and \eqref{eq:tau_N_phases}, which
ensure continuity across the transition. With this choice,
Eq.~\eqref{eq:tilde_A_T_transform} becomes
\begin{equation}
  \tilde{A}_T(\tau)=\sqrt{2k}\,A_T(\tau)
  \begin{cases}
    e^{i\tilde{k}H_{\rm end}\tau}, & \text{dS/qdS},\\[4pt]
    e^{-i\tilde{k}}, & \text{RD (from dS)},\\[4pt]
    e^{-i\tilde{k}/\alpha}, & \text{RD (from qdS)}.
  \end{cases}
\end{equation}

\subsection*{Equations of motion in terms of $N$}

Substituting Eq.~\eqref{eq:tilde_A_T_transform} into Eq.~\eqref{eq:transverse_eom}
gives, in the dS and qdS phases,
\begin{equation}
  \tilde{A}_T''(\tau)-2ik\,\tilde{A}_T'(\tau)+M^{2}a^{2}(\tau)\,\tilde{A}_T(\tau)=0,
  \label{eq:ATT_rescaled_cases}
\end{equation}
while in the RD epoch, after matching, the equation retains its original form,
\begin{equation}
  \tilde{A}_T''(\tau)+\left(k^{2}+M^{2}a^{2}(\tau)\right)\tilde{A}_T(\tau)=0 .
  \label{eq:transverse_mode_eom}
\end{equation}
Using Eq.~\eqref{eq:tau_to_N_derivatives}, both equations can be rewritten in terms
of the e-folding number $N$ as a single first-order system valid in all three regimes:
\begin{equation}
  \frac{d\tilde{A}_T}{dN}=\tilde{B}_T,\qquad
  \frac{d\tilde{B}_T}{dN}
  +\gamma\,\tilde{B}_T-2i\tilde{k}\,\mathcal{S}(N)\,\tilde{B}_T
  +\mathcal{D}(N)\,\tilde{A}_T=0 ,
  \label{eq:transverse_first_order_B}
\end{equation}
where $\gamma$ is defined in Eq.~\eqref{eq:gamma_definition} and
\begin{equation}
\mathcal{S}(N)=
\begin{cases}
e^{-N}, & \text{dS},\\
e^{-\alpha N}, & \text{qdS},\\
0, & \text{RD},
\end{cases}
\qquad
\mathcal{D}(N)=
\begin{cases}
\tilde{M}^{2}, & \text{dS},\\
\tilde{M}^{2}e^{2\epsilon N}, & \text{qdS},\\
\left(\tilde{k}^{2}+\tilde{M}^{2}e^{2N}\right)e^{2N}, & \text{RD}.
\end{cases}
\label{eq:SN_DN_definition}
\end{equation}
Here $\tilde{M}\equiv M/H_{\rm end}$, with $M$ given in
Eq.~\eqref{eq:Meff_piecewise}; explicitly, $\tilde M^2=\tilde m^2+12\xi$ (dS),
$\tilde M^2(N)=\tilde m^2+6\xi(1+\alpha)e^{-2\epsilon N}$ (qdS) and
$\tilde M^2=\tilde m^2$ (RD). Note that $1/\mathcal S^{2}=(aH/H_{\rm end})^{2}$
during inflation.

\subsection*{Initial conditions}

We impose second-order WKB Bunch--Davies initial conditions deep inside the
horizon at $N=N_{\rm ini}=-17$ in both the dS and qdS cases:
\begin{equation}
  \tilde A_T(N_{\rm ini})=\sqrt{\frac{\tilde k}{W(N_{\rm ini})}},
  \label{eq:initial_conditions_AT}
\end{equation}
\begin{equation}
  \tilde B_T(N_{\rm ini})=\tilde A_T(N_{\rm ini})
  \left[-\frac12\frac{W_N}{W}+i\bigl(\tilde k-W\bigr)\mathcal S(N)\right]_{N=N_{\rm ini}},
  \label{eq:initial_conditions_AT_derivative}
\end{equation}
where $W_N\equiv dW/dN$ and the arbitrary overall phase is chosen such that
$\tilde A_T(N_{\rm ini})$ is real. The second-order WKB frequency is
\begin{equation}
W^2(N)=\omega^2-\frac{\omega_{NN}+\gamma\,\omega_N}{2\,\mathcal S^2\,\omega}
+\frac{3}{4\,\mathcal S^2}\left(\frac{\omega_N}{\omega}\right)^2,
\qquad
\omega(N)=\sqrt{\tilde k^2+\tilde M^2(N)\,e^{2N}},
\label{eq:second_order_WKB_frequency}
\end{equation}
with $\omega_N=d\omega/dN$ and $\omega_{NN}=d^2\omega/dN^2$. The derivatives of
$\omega$ and $W$ are evaluated numerically using fourth-order central finite
differences. In the dS case $\gamma=1$ and $\mathcal S=e^{-N}$; in the qdS case
$\gamma=\alpha$ and $\mathcal S=e^{-\alpha N}$. For $\epsilon\to0$ the qdS
expressions reduce to the dS ones.

\subsection*{Matching to the RD epoch}

We assume an instantaneous transition from inflation to the RD epoch at $N=0$ and
impose continuity of the mode function and of its derivative, taking into account
the change of the phase in Eq.~\eqref{eq:tilde_A_T_transform}:
\begin{align}
\tilde{A}_T^{\rm RD}(N{=}0)&=\tilde{A}_T^{\rm dS/qdS}(N{=}0),
\label{eq:matching_AT}\\
\tilde{B}_T^{\rm RD}(N{=}0)&=\tilde{B}_T^{\rm dS/qdS}(N{=}0)-i\,\tilde k\,\tilde{A}_T(N{=}0).
\label{eq:matching_BT}
\end{align}
The resulting system is then evolved smoothly through the RD epoch.

\subsection*{Spectral energy density}

The physical derivative mode $B_T\equiv dA_T/dN$ is related to the rescaled
variables by $B_T\propto\tilde B_T-i\tilde k\,\mathcal S(N)\,\tilde A_T$, so that
\begin{align}
|A_T|^2&=\frac{|\tilde A_R|^2+|\tilde A_I|^2}{2H_{\rm end}\tilde k},
\label{eq:ABS_AT_rescaled}\\
|B_T|^2&=\frac{1}{2H_{\rm end}\tilde k}
\Bigl[\bigl(\tilde B_R+\tilde k\,\mathcal S\,\tilde A_I\bigr)^2
+\bigl(\tilde B_I-\tilde k\,\mathcal S\,\tilde A_R\bigr)^2\Bigr],
\label{eq:BT_squared}
\end{align}
where $\tilde A_{R,I}$ and $\tilde B_{R,I}$ denote the real and imaginary parts of
$\tilde A_T$ and $\tilde B_T$. The rescaled spectral energy density of the
transverse mode is
\begin{align}
a^3\frac{d\tilde{\rho}_{T}}{d\ln\tilde{k}}
={}& \frac{H_{\rm end}\tilde{k}^3e^{-N}}{2\pi^2}
\Bigg[e^{2\gamma N}|B_T|^2
\nonumber\\
&+\Bigl(\tilde{k}^2+\tilde{M}^2e^{2N}-6\xi(2+\gamma)e^{2\gamma N}\Bigr)|A_T|^2
\nonumber\\
&+12\xi\,e^{2\gamma N}\,\mathrm{Re}\left(B_TA_T^*\right)\Bigg].
\label{eq:spectral_rho_T_align}
\end{align}
We evaluate this quantity as a function of $N$, from the initial sub-horizon
regime, through horizon exit at $N=\ln\tilde k$ (dS) or $N=\ln\tilde k/\alpha$
(qdS), and into the RD era.

\clearpage
\section{Numerical Solutions for the Longitudinal Mode in dS, qdS, and RD Epochs}
\label{app:C}

In this appendix we describe the numerical solution of the longitudinal mode
during dS and qdS inflation, and its evolution
through the instantaneous transition to the RD epoch.
Tildes denote quantities in units of $H_{\rm end}$, i.e.\
$\tilde k=k/H_{\rm end}$, $\tilde m=m/H_{\rm end}$ and
$\tilde M=M/H_{\rm end}$, and the scale factor is normalized to
$a_{\rm end}=1$ at the end of inflation.

\noindent
The equation of motion for the longitudinal mode is
\begin{equation}
A_L''
+
\frac{2k^2}{k^2+M^2a^2}
\left(
\frac{M'}{M}+\frac{a'}{a}
\right)A_L'
+
\left(k^2+M^2a^2\right)A_L=0,
\label{eq:ALLL_eom}
\end{equation}
where a prime denotes differentiation with respect to conformal time $\tau$,
and the effective mass is
\begin{equation}
M^2(\tau)=m^2+\xi R(\tau).
\label{eq:M_eff_squared}
\end{equation}
In a spatially flat FLRW spacetime, the Ricci scalar is
\begin{equation}
R(\tau)=\frac{6a''}{a^3}.
\label{eq:ricci_scalar}
\end{equation}
We consider the dS, qdS, and RD backgrounds separately.

\subsection{dS regime}

In the dS limit, the Ricci scalar and the effective mass are
\begin{equation}
R=12H^2,
\qquad
M^2=m^2+12\xi H^2.
\label{eq:ricci_scalar_dS}
\end{equation}
Since $M$ is constant in this regime, Eq.~\eqref{eq:ALLL_eom} reduces to
\begin{equation}
A_L''
+
\frac{2k^2}{k^2+M^2a^2}
\left(\frac{a'}{a}\right)A_L'
+
\left(k^2+M^2a^2\right)A_L=0.
\label{eq:AL_dS_eom}
\end{equation}

\subsection{qdS regime}

For a qdS background, the Ricci scalar and the dimensionless effective mass are
\begin{equation}
R(\tau)=\frac{6}{a^2}\frac{1+\alpha}{\alpha^2\tau^2},
\qquad
\tilde M^2(\tau)
=
\tilde m^2
+
6\xi(1+\alpha)a^{-2\epsilon},
\label{eq:ricci_qds}
\end{equation}
where $\alpha=1-\epsilon$. Differentiating the effective mass gives
\begin{equation}
\frac{\tilde M'(\tau)}{\tilde M(\tau)}
=
-\frac{
6\epsilon\xi(1+\alpha)a^{-2\epsilon}
}{
\tilde m^2+
6\xi(1+\alpha)a^{-2\epsilon}
}
\left(\frac{a'}{a}\right).
\label{eq:Mprime_qds}
\end{equation}
Consequently,
\begin{equation}
\frac{\tilde M'(\tau)}{\tilde M(\tau)}
+\frac{a'}{a}
=
\left[
\frac{
\tilde m^2+
6\xi\alpha(1+\alpha)a^{-2\epsilon}
}{
\tilde m^2+
6\xi(1+\alpha)a^{-2\epsilon}
}
\right]
\left(\frac{a'}{a}\right).
\label{eq:M_a_prime_relation}
\end{equation}
The bracket is the ratio $G\equiv g/\tilde M^2$ that is defined in
Eq.~\eqref{eq:g_definition} below.

\subsection{RD regime}

During radiation domination, the Ricci scalar vanishes,
\begin{equation}
R=0,
\qquad
M^2=m^2.
\label{eq:ricci_RD}
\end{equation}
The longitudinal equation of motion therefore becomes
\begin{equation}
A_L''
+
\frac{2k^2}{k^2+m^2a^2}
\left(\frac{a'}{a}\right)A_L'
+
\left(k^2+m^2a^2\right)A_L=0.
\label{eq:EOM_L_RD}
\end{equation}
The initial data for the RD evolution are determined by matching the physical
longitudinal field and its canonical momentum at the inflation-to-radiation
transition; they are not imposed independently.

\subsection{Matching conditions and canonical normalization}

The longitudinal kinetic coefficient is
\begin{equation}
K_L(\tau)=\frac{M^2(\tau)a^2}{k^2+M^2(\tau)a^2}.
\label{eq:KL_definition}
\end{equation}
At the sudden transition, the physical matching conditions are
\begin{equation}
[A_L]_-^+=0,
\qquad
[K_LA_L']_-^+=0,
\label{eq:matching_conditions}
\end{equation}
where $[F]_-^+=F(0^+)-F(0^-)$. It is useful to introduce the canonical momentum
\begin{equation}
P_L\equiv K_LA_L'.
\label{eq:PL_definition}
\end{equation}
The matching conditions then take the simple form
\begin{equation}
A_L^+=A_L^-,
\qquad
P_L^+=P_L^-.
\label{eq:physical_matching}
\end{equation}
Since $K_L(\tau)$ can jump when the effective mass changes, neither $A_L'$ nor
the kinetic normalization is required to be continuous.

\noindent
The canonical field is
\begin{equation}
v_L=\sqrt{K_L}\,A_L.
\label{eq:canonical_longitudinal}
\end{equation}
In general, $\sqrt{K_L}$ is discontinuous at the transition. To express the
matching conditions in terms of $v_L$, define
\begin{equation}
q_\pm=
\left(\frac{d\ln\sqrt{K_L}}{dN}\right)_\pm,
\qquad
r=\frac{\sqrt{K_{L,+}}}{\sqrt{K_{L,-}}},
\qquad
N=\ln a.
\label{eq:matching_auxiliary}
\end{equation}
Using the continuity of $aH$ at the transition, the physical matching
conditions imply
\begin{equation}
v_L^+=r\,v_L^-,
\label{eq:canonical_matching_1}
\end{equation}
and
\begin{equation}
v_{L,N}^+
=
r^{-1}v_{L,N}^-
+
\left(rq_+-r^{-1}q_-\right)v_L^-.
\label{eq:canonical_matching_2}
\end{equation}
Thus, continuity of the physical field and its canonical momentum does not
imply continuity of $v_L$ or its $N$-derivative.

\noindent
For numerical evolution, an alternative is to evolve $A_L$ and $P_L$ directly.
In terms of $N$, their first-order equations are
\begin{equation}
\frac{dA_L}{dN}
=
\frac{P_L}{aHK_L},
\qquad
\frac{dP_L}{dN}
=
-\frac{aM^2}{H}A_L.
\label{eq:AL_PL_evolution}
\end{equation}
This formulation implements the physical matching conditions directly and
avoids differentiating the discontinuous canonical normalization $\sqrt{K_L}$
across the transition.

\noindent
If a phase-rotated numerical variable is introduced, its phase factor must be
specified independently of $\sqrt{K_L}$. In particular, a continuous phase
convention does not require continuity of the kinetic normalization.

\subsection{Canonical equation in terms of the e-folding number}

The canonical longitudinal variable is
\begin{equation}
V
=
\sqrt{\frac{M^2a^2}{k^2+M^2a^2}}\,A_L .
\label{eq:V_definition}
\end{equation}

\noindent
Using Eq.~\eqref{eq:ALLL_eom}, the canonical longitudinal variable satisfies
the following equation in conformal time:
\begin{equation}
\begin{aligned}
V''
&+
\frac{-\tilde{k}^4+
2\tilde{k}^2\tilde{M}^2a^2}
{\left(\tilde{k}^2+\tilde{M}^2a^2\right)^2}
S_L^2(N(\tau))V
\\
&-
\frac{\tilde{k}^2}
{\tilde{k}^2+\tilde{M}^2a^2}
S_L'(N(\tau))V
\\
&+
H_{\rm end}^2
\left(\tilde{k}^2+\tilde{M}^2a^2\right)V=0,
\end{aligned}
\label{eq:VLL_eom}
\end{equation}
where $S_L(N)$ is defined in Eq.~\eqref{eq:S_definition}. Using the conversion
between conformal time and the e-folding number, this equation becomes
\begin{equation}
\begin{aligned}
&
\frac{d^2V}{dN^2}
+\gamma\frac{dV}{dN}
\\
&+
\frac{-\tilde{k}^4+
2\tilde{k}^2\tilde{M}^2a^2}
{\left(\tilde{k}^2+\tilde{M}^2a^2\right)^2}
\frac{
\left[
\tilde{m}^2a^2+
6\xi\gamma(1+\gamma)a^{2\gamma}
\right]^2
}{
\tilde{M}^4a^4
}V
\\
&-
\frac{\tilde{k}^2}
{\tilde{k}^2+\tilde{M}^2a^2}
\Bigg[
\gamma
-
\frac{
72\epsilon^2\xi^2(1+\gamma)^2a^{4\gamma}
}{
\tilde{M}^4a^4
}
\\
&\hspace{4em}
+
\frac{
6\epsilon\xi(3\epsilon-1)(1+\gamma)a^{2\gamma}
}{
\tilde{M}^2a^2
}
\Bigg]V
\\
&+
\left(\tilde{k}^2+\tilde{M}^2a^2\right)
e^{-2\gamma N}V=0.
\end{aligned}
\label{eq:EOM_L_eff}
\end{equation}
Here $\gamma$ is defined in Eq.~\eqref{eq:gamma_definition}. In the RD regime,
where $\gamma=-1$ and $\tilde M=\tilde m$, this equation reduces to
\begin{equation}
\frac{d^2V}{dN^2}
-\frac{dV}{dN}
+
\frac{
3\tilde{m}^2e^{2N}\tilde{k}^2
}{
\left(\tilde{k}^2+\tilde{m}^2e^{2N}\right)^2
}V
+
\left(\tilde{k}^2+\tilde{m}^2e^{2N}\right)
e^{2N}V=0.
\label{eq:longitudinal_canonical_eom}
\end{equation}

\subsection{Phase convention and reconstruction of the longitudinal mode}

We evolve a phase-rotated numerical variable $\tilde A_L$, related to the
canonical variable by
\begin{equation}
V=
\frac{
e^{i\tilde{k}e^{-\alpha N}/\alpha}
}{
\sqrt{2H_{\rm end}\tilde{k}}
}
\,\tilde{A}_L ,
\label{eq:V_to_AL}
\end{equation}
The physical longitudinal mode function is related to the
phase-rotated variable by $A_L=P\,\tilde A_L$, where
\begin{equation}
P(N)=
\sqrt{
\frac{\tilde{k}^2+\tilde{M}^2e^{2\gamma N}}
{2\tilde{k}\,\tilde{M}^2e^{2\gamma N}}
}.
\label{eq:P_definition}
\end{equation}
Here, $\gamma$ is defined in Eq.~\eqref{eq:gamma_definition}.
The phase convention must be consistent with the evolution equation
in each background. Accordingly, the coefficient entering the
phase-derivative terms is defined piecewise as
\begin{equation}
\Omega(N)=
\begin{cases}
e^{-N}, & \mathrm{dS},\\[2pt]
e^{-\alpha N}, & \mathrm{qdS},\\[2pt]
0, & \mathrm{RD}.
\end{cases}
\label{eq:phase_coefficient}
\end{equation}
In the RD regime, the phase is held constant at its value at the transition,
so that its derivative vanishes and $\Omega(N)=0$. A running phase in RD would
instead require its corresponding rotating-frame evolution equation.

\noindent
The real and imaginary components of the numerical variable are used to
reconstruct the physical longitudinal quantities. Writing
\begin{equation}
\tilde{A}_L=\tilde{A}_R+i\tilde{A}_I,
\qquad
\tilde{B}_L=\frac{d\tilde{A}_L}{dN}
=\tilde{B}_R+i\tilde{B}_I,
\label{eq:complexx_longitudinal_variables}
\end{equation}
we define
\begin{equation}
\begin{aligned}
\Phi_L={}&
\tilde{B}_R
+\tilde{k}\Omega(N)\tilde{A}_I
-
\frac{\tilde{k}^2}
{\tilde{k}^2+\tilde{M}^2e^{2N}}
\frac{g(N)}{\tilde{M}^2}\tilde{A}_R,
\\
\Psi_L={}&
\tilde{B}_I
-\tilde{k}\Omega(N)\tilde{A}_R
-
\frac{\tilde{k}^2}
{\tilde{k}^2+\tilde{M}^2e^{2N}}
\frac{g(N)}{\tilde{M}^2}\tilde{A}_I.
\end{aligned}
\label{eq:Phi_Psi_LL}
\end{equation}
The background-dependent function $g(N)$ is
\begin{equation}
g(N)=
\begin{cases}
\tilde m^2+12\xi,
& \mathrm{dS},
\\[4pt]
\tilde m^2+6\xi\alpha(1+\alpha)e^{-2\epsilon N},
& \mathrm{qdS},
\\[4pt]
\tilde m^2,
& \mathrm{RD},
\end{cases}
\label{eq:g_definition}
\end{equation}
and we write $G(N)\equiv g(N)/\tilde M^2(N)$ for the dimensionless ratio. The
piecewise coefficient $\Omega(N)$ is used in Eq.~\eqref{eq:Phi_Psi_LL} when
the phase is frozen during RD.

\noindent
The reconstructed quadratic quantities are
\begin{equation}
|A_L|^2
=
\frac{
\tilde{k}^2+\tilde{M}^2e^{2N}
}{
2H_{\rm end}\tilde{k}\tilde{M}^2e^{2N}
}
\left(\tilde{A}_R^2+\tilde{A}_I^2\right),
\label{eq:AL_squared}
\end{equation}
and
\begin{equation}
|B_L|^2
=
\frac{
\tilde{k}^2+\tilde{M}^2e^{2N}
}{
2H_{\rm end}\tilde{k}\tilde{M}^2e^{2N}
}
\left(\Phi_L^2+\Psi_L^2\right),
\label{eq:BL_squared}
\end{equation}
where $B_L\equiv dA_L/dN$.

\subsection{Initial conditions}
\label{sec:IC_L}

We impose second-order WKB Bunch--Davies initial conditions deep inside the
horizon at $N=N_{\rm ini}=-17$ in both the dS and qdS cases. Working in units
$H_{\rm end}=1$, with $a=e^{N}$ and $aH/H_{\rm end}=e^{\gamma N}$, we define
\begin{equation}
\begin{aligned}
C(N) &= 6\xi(1+\gamma)\,e^{-2\epsilon N}, & \qquad
\tilde M^2(N) &= \tilde m^2 + C(N), \\
\Delta(N) &= \tilde k^2 + \tilde M^2 e^{2N}, & \qquad
c(N) &= \frac{\tilde k^2}{\Delta(N)}.
\end{aligned}
\end{equation}
and
\begin{equation}
G(N)\equiv\frac{g(N)}{\tilde M^2}=\frac{\tilde m^2+\gamma\,C(N)}{\tilde M^2(N)},
\qquad
G_{,N}=\frac{2\epsilon^2\,\tilde m^2\,C(N)}{\tilde M^4(N)} .
\end{equation}
Here $\gamma=1$ and $\epsilon=0$ in the dS case, and $\gamma=\alpha=1-\epsilon$
in the qdS case, so that $G=1$ and $G_{,N}=0$ in dS. The logarithmic derivative
of the kinetic normalization, $s\equiv d\ln\sqrt{K_L}/dN$, is
\begin{equation}
s=c\,G,\qquad
s_{,N}=-2c(1-c)\,G^2+c\,G_{,N}.
\end{equation}
The canonical frequency of $V=\sqrt{K_L}\,A_L$ is
\begin{equation}
\omega^2(N)=\Delta(N)-e^{2\gamma N}\bigl(s_{,N}+s^2+\gamma\,s\bigr),
\end{equation}
and the second-order WKB frequency is
\begin{equation}
W^2(N)=\omega^2
-\frac{e^{2\gamma N}}{2\omega}\bigl(\omega_{,NN}+\gamma\,\omega_{,N}\bigr)
+\frac34\,e^{2\gamma N}\frac{\omega_{,N}^{2}}{\omega^{2}},
\end{equation}
where $\omega_{,N}\equiv d\omega/dN$, and the derivatives of $\omega$ and $W$
are evaluated numerically with fourth-order central finite differences. The
numerical variable of Eq.~\eqref{eq:V_to_AL} is then initialized as
\begin{equation}
\tilde A_L(N_{\rm ini})=\sqrt{\frac{\tilde k}{W(N_{\rm ini})}},\qquad
\frac{d\tilde A_L}{dN}\bigg|_{N_{\rm ini}}=
\left[-\frac{W_{,N}}{2W}+i\,(\tilde k-W)\,e^{-\gamma N}\right]_{N_{\rm ini}}
\tilde A_L(N_{\rm ini}),
\end{equation}
where the arbitrary overall phase is chosen so that $\tilde A_L(N_{\rm ini})$
is real. In terms of the physical mode function $A_L=P\,\tilde A_L$, this gives
\begin{equation}
\begin{aligned}
A_L(N_{\rm ini})&=P(N_{\rm ini})\sqrt{\frac{\tilde k}{W(N_{\rm ini})}},\\[4pt]
\frac{dA_L}{dN}\bigg|_{N_{\rm ini}}&=
\left[-\frac{W_{,N}}{2W}-c\,G-i\,W\,e^{-\gamma N}\right]_{N_{\rm ini}}A_L(N_{\rm ini}).
\end{aligned}
\end{equation}
For $\tilde k/(aH)\gg1$ these reduce to
$|A_L|\to\sqrt{\tilde k/2}\,/(aM)$ and $dA_L/dN\to-i\,(k/aH)\,A_L$. For
$\epsilon\to0$ the qdS expressions reduce exactly to the dS ones.

\subsection{Spectral energy density of the longitudinal mode}

The bare spectral energy density of the longitudinal mode, including the
nonminimal contributions to the Hilbert stress-energy tensor, can be expressed
as
\begin{equation}
\begin{aligned}
a^3
\frac{d\tilde{\rho}_L}{d\ln\tilde{k}}
={}&
\frac{H_{\rm end}\tilde{k}^3e^{-N}}{4\pi^2}
\Bigg[
\frac{
\tilde{M}^2e^{2(1+\gamma)N}
}{
\tilde{k}^2+\tilde{M}^2e^{2N}
}
|B_L|^2
+
\tilde{M}^2e^{2N}|A_L|^2
\\
&+
\xi\Bigg\{
12e^{2\gamma N}
\left(
1+
\frac{\tilde{k}^2}
{\tilde{k}^2+\tilde{M}^2e^{2N}}
\right)
\operatorname{Re}(B_LA_L^*)
\\
&\qquad+
\frac{
24\tilde{k}^2e^{4\gamma N}
}{
\left(\tilde{k}^2+\tilde{M}^2e^{2N}\right)^2
}
\,G(N)\,
|B_L|^2
\\
&\qquad+
6(2+\gamma)e^{2\gamma N}
\left[
\frac{
\tilde{k}^2e^{2\gamma N}
}{
\left(\tilde{k}^2+\tilde{M}^2e^{2N}\right)^2
}
|B_L|^2
-|A_L|^2
\right]
\Bigg\}
\Bigg].
\end{aligned}
\label{eq:rhoLL_eff}
\end{equation}
Here $\gamma$ is defined in Eq.~\eqref{eq:gamma_definition}, and the
appropriate effective mass is used in each background; in particular,
$\tilde M=\tilde m$ during RD. The quantities in Eq.~\eqref{eq:rhoLL_eff} are
bare spectral energy contributions obtained from the Hilbert stress-energy
tensor. They should not be identified with a renormalized physical energy
density without specifying and implementing an appropriate renormalization
prescription.

\clearpage
\section{Diagnostic of the Sharp-Transition Limit}
\label{app:D}

In this appendix, we discuss the implications of treating the transition from
inflation to radiation domination as instantaneous. We restrict the discussion
to the longitudinal mode, which is the sector evolved in the numerical analysis
of Appendix~\ref{app:C}. The longitudinal mode is evolved in dS,
qdS and RD backgrounds, with the
inflationary and RD solutions connected through matching conditions at $N=0$.
The transition is therefore treated in the sharp-transition limit rather than
through a smooth interpolation of the background.

\noindent
We take the scale factor and the Hubble parameter to be continuous across the
transition,
\begin{equation}
[a]_0=0,
\qquad
[H]_0=0,
\label{eq:D_background_matching}
\end{equation}
where the jump of any quantity $Q$ is defined as
\begin{equation}
[Q]_0\equiv Q(0^+)-Q(0^-).
\label{eq:D_jump_definition}
\end{equation}
The matching conditions for the longitudinal mode are imposed on the physical
field and its canonical momentum,
\begin{equation}
[A_L]_0=0,
\qquad
[K_L A_L']_0=0,
\label{eq:D_longitudinal_matching}
\end{equation}
where $K_L$ is the longitudinal kinetic coefficient defined in
Eq.~\eqref{eq:KL_definition}. Since $aH$ is continuous, the second condition is
equivalent to continuity of $K_L\,dA_L/dN$. Because the effective mass changes
at the transition, $K_L$ can jump, and neither $A_L'$ nor the kinetic
normalization is required to be continuous.

\subsection{Radiation-Dominated Background}

After the transition at $N=0$, the Universe enters a RD phase,
with
\begin{equation}
\frac{H}{H_{\rm end}}=e^{-2N}.
\label{eq:D_RD_Hubble}
\end{equation}
For a spatially flat FLRW background, $R=0$, so
\begin{equation}
M^2=m^2+\xi R=m^2.
\label{eq:D_RD_effective_mass}
\end{equation}
Nevertheless, the non-minimal contribution to the stress tensor must be
retained, since it includes curvature and derivative terms that need not vanish
when $R=0$.

\subsection{Longitudinal Sector and the Transition}

The non-dynamical component $A_0$ is determined by a constraint that depends on
the effective mass and the background. Its one-sided limits may therefore
differ across the inflation-to-radiation transition, even when the longitudinal
matching conditions are imposed. Consequently,
\begin{equation}
X=A_\mu A^\mu
\label{eq:D_X_definition}
\end{equation}
may also be discontinuous. Since the non-minimal stress tensor contains
derivatives of $X$, distributional contributions may arise in the
sharp-transition limit.

\noindent
The matching conditions in Eq.~\eqref{eq:D_longitudinal_matching} do not by
themselves determine these contributions. We therefore interpret the
longitudinal energy density obtained from the regular solutions as a one-sided
quantity, and we assign no unique value to the full energy density exactly at
$N=0$.

\subsection{Radiation-Era Evolution and Late-Time Oscillations}

The RD solution is evolved from the matched initial conditions at $N=0$,
describing the regular longitudinal mode away from the instantaneous
transition. With $a_{\rm end}=H_{\rm end}=1$, we choose the conformal time as
\begin{equation}
\tau=e^N=a.
\label{eq:D_RD_conformal_time}
\end{equation}
The canonical longitudinal variable $V$ satisfies
\begin{equation}
\frac{d^2V}{dN^2}-\frac{dV}{dN}+\Omega_N^2\,V=0,
\label{eq:D_RD_canonical_equation}
\end{equation}
where
\begin{equation}
\Omega_N^2=
\frac{3\tilde m^2a^2\tilde k^2}{(\tilde k^2+\tilde m^2a^2)^2}
+a^2\left(\tilde k^2+\tilde m^2a^2\right),
\label{eq:D_RD_effective_frequency}
\end{equation}
with $\tilde m=m/H_{\rm end}$ and $\tilde k=k/H_{\rm end}$. This is
Eq.~\eqref{eq:longitudinal_canonical_eom}, written with $\Omega_N$ for the
effective frequency.

\noindent
At sufficiently late times, when $k\ll ma$ and the mass term dominates, the
equation reduces to
\begin{equation}
\frac{d^2V}{d\tau^2}+m^2\tau^2\,V\simeq0.
\label{eq:D_late_time_oscillator}
\end{equation}
The WKB phase is $\theta\simeq m\tau^2/2$, so an energy-density contribution
oscillating with phase $2\theta\simeq m\,e^{2N}$ has the local e-fold period
\begin{equation}
\Delta N\simeq\frac{\pi}{m\,e^{2N}}.
\label{eq:D_oscillation_period}
\end{equation}
For $\tilde m=10^{-2}$ at $N=4.5$, this gives $\Delta N\simeq0.0388$.

\noindent
Finally, the radiation-era condition $H=m$ occurs at
\begin{equation}
N_{H=m}=\frac12\ln\!\left(\frac1{\tilde m}\right)\simeq2.30259
\qquad(\tilde m=10^{-2}).
\label{eq:D_H_mass_crossing}
\end{equation}
This background-dependent condition is distinct from the momentum-dependent
crossing $k=aM$.

\subsection{Interpretation of the Numerical Results}

The numerical solutions describe the longitudinal mode evolved through the
inflationary and RD phases, with the two solutions connected
by the matching conditions at $N=0$. The RD curves represent the regular
one-sided solutions after the transition.

\noindent
In particular, the longitudinal energy-density spectra shown in
Figs.~\ref{fig:rho_L_Total} and~\ref{fig:rho_L_Total_sup} should be interpreted
as the evolution of the regular longitudinal contribution on either side of the
sharp transition at $N=0$. The two branches are plotted separately, and no
interpolation is made across $N=0$.

\noindent
The sharp-transition treatment is therefore a diagnostic approximation that
allows the longitudinal mode evolution before and after the transition to be
studied separately. A treatment of the transition over a finite time interval
would be required to determine the detailed evolution of the energy-momentum
tensor through the transition itself.

\section*{Acknowledgements}

E.~M. and F.~S. thank Oleg Lebedev, Jiro Soda, and Ying-Li Zhang for valuable comments and helpful discussions, which improved the presentation and clarity of this work. The work of H.~K. is supported in part by the National Natural Science Foundation of China (NSFC) under Grant Nos.~W2533015, 12375052, and 11935009, and by the Tianjin University Self-Innovation Fund for Extreme Basic Research under Grant No.~2025XJ21-0007.

\nocite{*}

\bibliographystyle{unsrtnat}
\bibliography{references}

\end{document}